\documentclass[iop]{emulateapj}

\newcommand{\hone}{H~{\footnotesize{I}}}  	
\newcommand{\carfour}{C~{\footnotesize{IV}}}  	
\newcommand{\nitfive}{N~{\footnotesize{V}}}  	
\newcommand{\oxysix}{O~{\footnotesize{VI}}}  	
\newcommand{\oxyseven}{O~{\footnotesize{VII}}}  
\newcommand{\oxyeight}{O~{\footnotesize{VIII}}} 
\newcommand{\chandra}{{\it{Chandra}}}		
\newcommand{\rosat}{{\it{ROSAT}}}		
\newcommand{\xmmnewton}{{\it{XMM-Newton}}}	
\newcommand{\warmism}{C}			
\newcommand{\purehd}{B}				
\newcommand{\hot}{A}				
\newcommand{\nei}{Br}				
\newcommand{\adiabatic}{Ba}			

\begin{document}
\title{Modeling the X-rays Resulting From High Velocity Clouds}

\shorttitle{X-rays from HVCs}
\shortauthors{Shelton et al.}


\author{R. L. Shelton\altaffilmark{1},
	K. Kwak\altaffilmark{1,2},
	and D. B. Henley\altaffilmark{1}}
\affil{$^1$Department of Physics and Astronomy and the Center
for Simulational Physics, the University of Georgia, Athens, GA 30602;
rls@physast.uga.edu, kkwak@physast.uga.edu, dbh@physast.uga.edu\\
$^2$Radio Astronomy Research Center, Korea Astronomy and Space Science Institute, 61-1 Hwaam-dong, Yuseong-gu, Dajeon 305-348, South Korea;\\
kkwak@kasi.re.kr}

\begin{abstract}

With the goal of understanding why X-rays have been reported near some 
high velocity clouds, we perform detailed 3 dimensional hydrodynamic and 
magnetohydrodynamic simulations of clouds interacting  with environmental 
gas like that in the Galaxy's thick disk/halo or the Magellanic Stream. 
We examine 2 scenarios.  In the first, clouds travel fast enough to 
shock-heat warm environmental gas.    In this scenario, the X-ray 
productivity depends strongly on the speed of the cloud and the radiative 
cooling rate.  In order to shock-heat environmental gas to temperatures of 
$\geq 10^6$~K, cloud speeds of $\geq 300$~km/s are required.  
If cooling is quenched, then the  shock-heated 
ambient gas is X-ray emissive, producing bright  X-rays in the 1/4 keV 
band and some X-rays in the 3/4 keV band due to \oxyseven\ and other ions.  
If, in contrast, the radiative cooling rate  is similar to that of 
collisional ionizational equilibrium plasma with solar abundances, 
then the shocked gas is only mildly bright and for only about 1 Myr.  
The predicted count rates for the non-radiative case are bright enough 
to explain the count rate observed with \xmmnewton\ toward a Magellanic 
Stream cloud and some enhancement in the \rosat\ 1/4~keV count rate toward 
Complex C, while the predicted count rates for the fully radiative case 
are not.  In the second scenario, the clouds travel through and mix with 
hot ambient gas.  The mixed zone can contain hot gas, but the hot 
portion of the mixed gas is not as bright as  those from the shock-heating 
scenario. \\
\vspace{0.0cm}
\end{abstract}

\section{Introduction}
\label{sect:intro}

Observations of diffuse gas have found a population of
massive, fast moving clouds.
With line-of-sight speeds between $\sim90$ km s$^{-1}$ 
and $\sim300$ km s$^{-1}$ 
\citep{wakker_vanwoerden_91},
they are appropriately named high velocity clouds (HVCs).
These clouds are plentiful;
neutral HVC material
covers $37\%$ of the sky \citep{murphy_etal_95}
while intermediately and highly ionized HVC material covers
$\sim 80\%$ and
$\gtrsim60\%$ 
of the sky, respectively
\citep{shull_etal_09,sembach_etal_03}.
Although some of the high velocity material is isolated, many
of the clouds are grouped into large complexes.
Complex C, for example, 
stretches from 
$\ell \sim 30\degr$ to $\ell \sim 140\degr$, 
while the
Magellanic Stream 
runs along $\ell = 90\degr$ from 
$b \sim -40\degr$ to near the South Galactic Pole and then resumes
again, running along $\ell = 290\degr$ from the South
Galactic Pole to $b \sim -30\degr$. 

Several clouds are relatively near to the Galactic plane.   
The Smith Cloud (also called Complex GCP), Complex A,
the Cohen Stream, which is part of the Anticenter Complexes, 
and Complex C are between
$2.5$ and $4$~kpc,
2.5 and 7~kpc,
$4$ and $9$~kpc,
and 5 and 10~kpc, respectively, above or below the Galactic midplane 
\citep{wakker_etal_07,wakker_etal_08,lockman_etal_08,vanwoerden_etal_99,thom_etal_08}.
These clouds are near enough to the plane to be
interacting with the Galaxy's thick disk/halo
(see \citet{santillan_etal_99} and references within).

\begin{figure}
\epsscale{1.2}
\plotone{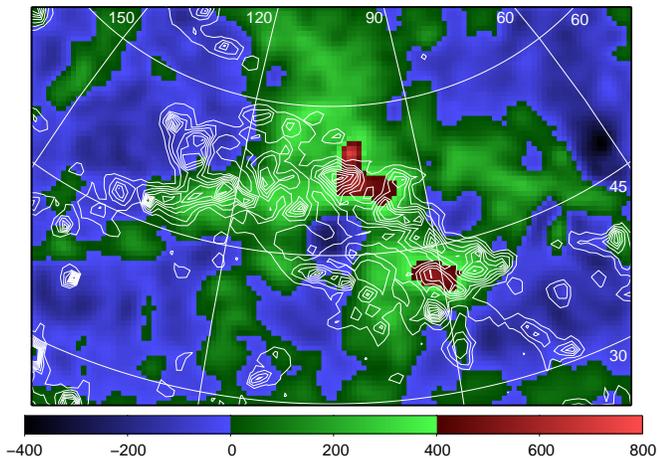}
  \caption{
Excess X-ray count rate in the region of Complex C.
The excess is denoted by color,
in units of $10^{-6}$ \rosat\ 1/4 keV counts~s$^{-1}$~arcmin$^{-2}$, and 
was calculated by subtracting 
a model which included smooth foreground and 
distant components from the observed \rosat\ All Sky Survey count rate.
The distant component of our subtracted
background model experienced extinction by Galactic \hone.
Note that the plotted values are the observed excess count rates;
they have not been de-absorbed.
The white contours show the location of \hone\ 
with velocity, $v < -90$~km~s$^{-1}$.   
The contour spacing is $10^{19}$ cm$^{-2}$.
Latitude is marked by the upward curving lines.   They
are spaced 15$^{\rm{o}}$ apart, with the lowest visible line
marking $b = 30^{\rm{o}}$.    Longitude is marked by 
radiating lines.   They are spaced 30$^{\rm{o}}$ apart,
with the left-most visible line marking $l = 150^{\rm{o}}$.
The map shows that the brightest excess 1/4~keV count rates are
aligned with Complex C and that the Complex C region is
generally bright in soft X-rays.
\\
}
\label{fig:complexcmap}
\end{figure}

Cloud-ISM interactions may be identified through 
anticorrelations with Galactic \hone\ at normal velocities
\citep{morras_etal_98}, 
high ion ratios \citep{tripp_etal_03}, 
and
X-ray enhancements (e.g., \citealt{hirth_etal_85}).
For example, \citet{hirth_etal_85}
noted an excess soft X-ray surface brightness
near high velocity \hone\ gas in Draco, a region 
now considered the southern part of Complex C.
Later, 
\citet{herbstmeier_etal_95}
reported excess 1/4~keV X-rays on the edge of Complex M,
\citet{kerp_etal_96}
noted excess 1/4~keV X-rays
in the Complex C region, and
\citet{kerp_etal_99} reported 1/4~keV excesses for 
Complexes C, D, and GCN.
(See Figure~\ref{fig:complexcmap} for our estimate of the Complex C excess.)
Generally, diffuse 1/4~keV X-rays are interpreted as tracers of 
$\sim 10^6$~K gas.  
A slight excess of somewhat more energetic X-rays 
has been reported for a sight line
through the Magellanic Stream
\citep{bregman_etal_09}.  
Although the clouds in the Magellanic Stream are not interacting with 
the Milky Way's disk, they may be interacting with the 
extended halo or with gas ablated from preceding clouds in the
stream.

Like \citet{hirth_etal_85}, \citet{bregman_etal_09} 
suggested that the excess X-rays
may have been emitted by shock-heated gas.
Shock heating would be 
possible if the collision speed were multi-hundred km~s$^{-1}$
and the gas were initially warm or hot.
For unmagnetized warm plasmas, the post shock temperature
would be $\sim 1 \times 10^6$~K if the collision speed were
$300$~km~s$^{-1}$
\citep{shu_92}.
Such a speed may be achieved by Magellanic Stream clouds, given
that the orbital velocity of the Magellanic Stream is 
$378 \pm 18$~km~s$^{-1}$ \citep{blandhawthorn_etal_07}.
As is the case for most HVCs, the impact speed of Complex C is unknown.
Most HVCs have line-of-sight velocities $\la300$~km~s$^{-1}$, but
their total velocities may be much larger than their line-of-sight
velocities if the angles between
the HVCs' velocities and the lines-of-sight are large.
This is the case for the Smith Cloud, whose total velocity has been
calculated from the variation in line-of-sight velocity as a function
of observing angle and other observables 
to be $\sim300$~km~s$^{-1}$, while its
line-of-sight velocity is only
$\sim100$~km~s$^{-1}$ \citep{lockman_etal_08}.
In addition, an HVC's currently observed velocity may be
less than its velocity when it first encountered the Galaxy.
We examine the X-ray productivity of gas shocked by
fast clouds in this paper.

Turbulent mixing should also be considered.
As the cloud passes through the ambient medium,
shear instabilities develop at the contact surface.  Gas on either
side of the interface mixes, resulting in a zone of intermediate
temperature, intermediate density gas.   This logic has been used
to explain high ions associated with HVCs 
\citep{tripp_etal_03,fox_etal_04,kwak_etal_11}.   
In this paper, we also consider
the possibility that some of the transition zone gas may be 
hot enough and dense enough to yield observable quantities of
X-rays.   

Other potential mechanisms include magnetic reconnection
\citep{kerp_etal_94,zimmer_etal_96,zimmer_etal_97,kerp_etal_99}. 
As high velocity clouds move through the halo and thick
disk, they should deform and compress the magnetic field.   
\citet{zimmer_etal_96,zimmer_etal_97} suggest that 
shear between
the cloud and the ambient plasma will turbulently mix
the magnetic field in the region very near to the cloud.  
When these magnetic field 
lines reconnect, they release energy.
\citet{zimmer_etal_96,zimmer_etal_97} performed analytic
and resistive magnetohydrodynamic simulations of the system
finding that magnetic reconnections can release enough
energy to heat the gas to $>10^6$~K.
Because the magnetic reconnection scenario has already been 
examined
with magnetohydrodynamic simulations, it
is not simulated again in this paper.
Recently, additional ideas have been suggested.
Noting the X-rays emission that follows after charge exchange 
reactions in the heliosphere,
\citet{provornikova_etal_11} suggest that charge exchange
may be important at the interfaces between clouds and hot
gas, and,
noting the possible role of MHD plasma waves 
in heating the solar corona \citet{jelinek_hensler_11} 
suggest that plasma waves 
instigated by collisions between
high velocity clouds and halo gas 
may also be important.


In order to better understand HVCs, their interactions with
the Galaxy, and the possibility that they may induce X-rays,
we perform FLASH magnetohydrodynamic 
simulations of the shock heating and turbulent mixing scenarios.    
Our simulations begin with a cloud of similar size and
\hone\ column density as the lumps
in Complex C.
%
In our simulations of 
the shock heating scenario, the cloud initially has a speed of
$\geq 200$~km~s$^{-1}$
relative to the ambient gas.
We examine the effect of ambient density, using
moderate (e.g., $n \sim 7 \times 10^{-3}$ atoms~cm$^{-3}$) and 
lower densities for the ambient gas.
The moderately dense ambient medium 
may represent material ablated from a preceding HVC or a thick
disk/halo cloud.
We found that when $\geq300$~km~s$^{-1}$ clouds interact with
moderate density ambient media and radiative cooling is disabled,
the shocked, compressed ambient medium yields extremely bright X-rays for $\sim10$~Myr.
If, in contrast, radiative cooling proceeds at the 
collisional ionizational equilibrium (CIE) rate, then the gas is 
only moderately bright and for only $\sim1$~Myr.
In both the adiabatic and the radiatively cooling simulations,
greater ambient densities resulted in greater emission intensities.
In our turbulent mixing simulations, a cool
cloud falls through hot 
halo gas that is in hydrostatic equilibrium.
In the simulations having very hot ambient
media ($T = 3 \times 10^6$~K), 
the mixed zone contains some $T \sim 10^6$~K gas, which is hot
enough to produce 1/4 keV X-rays in CIE calculations.
However, the mixed gas falls behind
the cloud, into a region where the pressure and density are low.   It is the 
low density that limits the X-ray production of this scenario, 
causing the surface brightnesses to be too small to be detectable.

In Section~\ref{sect:technique}, we describe the modeling algorithms and list
the simulational parameters.    Section~\ref{sect:results} presents
the results.
In Section~\ref{subsect:cievsnei}, we compare CIE and non-equilibrium 
ionization (NEI)
calculations, finding that the CIE approximation yields similar
1/4 keV X-ray spectra to the NEI calculations.
In Section~\ref{subsect:shockheating}, we evaluate the ability of fast HVC collisions to shock heat the
ambient gas and induce X-ray emission.  
The predictions include 
1/4 keV surface brightnesses, \oxyseven\ and \oxyeight\ intensities,
and \oxyseven\ column densities.
In Section~\ref{subsect:hotism}, we evaluate the ability of turbulent mixing
to create hot, X-ray emissive gas.
Section~\ref{subsect:obsappearance} 
describes how the simulated clouds would appear to observers
while
Section~\ref{subsect:resolution} discusses higher and lower resolution 
simulations, concluding that the numerical resolutions used in 
our earlier simulations are adequate.
The results are summarized in Section~\ref{sect:discussion}.
\\

\section{Modeling Technique}
\label{sect:technique}

\subsection{Magnetohydrodynamic Algorithms}
\label{subsect:mhd}
We use similar hydrodynamic and magnetohydrodynamic
algorithms as in \citet{kwak_etal_09,kwak_shelton_10}.
To wit, we use the FLASH computer code, version 2.5 
\citep{fryxell_etal_00}
with adaptive mesh refinement (AMR)
to model the hydrodynamics and magnetohydrodynamics of 
fast moving clouds and ambient gas.   
For our suite of cloud shock simulations 
(Section~\ref{subsect:shockheating}),
we test the effects of radiative cooling by modeling it in some
simulations but not others.
The radiative cooling calculations are done with the
FLASH module, which uses cooling rates for CIE plasmas.
We do not simulate gravity in our cloud shock simulations,
but instead start the cloud with a large initial velocity.
We do the same in our CIE testing (Section~\ref{subsect:cievsnei})
and resolution experiments (Section~\ref{subsect:resolution}).
In our turbulent mixing simulations (Section~\ref{subsect:hotism})
we use the gravity module in FLASH and the
expression for gravitational acceleration 
presented in \citet{ferriere_98} to
establish hydrostatic equilibrium in the background gas and to
simulate the Milky Way's gravitational pull on the cloud.
In order to maintain hydrostatic balance in the background
gas, radiative cooling must be disallowed and is so 
in the turbulent mixing simulations.
It is not needed in the CIE testing 
and so is disabled in them as well.
%
In two of our simulations, we model 
a magnetic field oriented perpendicular to the cloud's motion.
The others have no magnetic field.
For the purposes of the magnetohydrodynamics, we treat 
all of the gas as if it is fully ionized in all of our 
simulations.  Thermal conduction is not modeled.

The predictions of the X-ray count rates and very high ion
column densities require predictions of the fractions of
ions in any given state in the gas.
We calculate the ion fractions using collisional ionizational 
equilibrium (CIE) and/or
partially non-equilibrium ionization (NEI) 
models.
In the CIE models, the ionization levels are calculated from the
temperature of the simulated gas, using the
Raymond and Smith code (\citet{raymond_smith_77}, with
updates).   For the NEI models, we enable FLASH to track the
ionization levels of select elements as a function of time
by accounting for their collisional ionization and 
recombination during each timestep.
Ionization and recombination rates from \citet{summers_74}
are used.   We limit our NEI calculations to only two
elements, silicon and oxygen, in order to economize on CPU and
memory.   
Silicon is a major contributer to X-ray spectra in the
1/4 keV band, while oxygen is an important contributer in the 3/4~keV band.
In Section~\ref{subsect:cievsnei}, we show that CIE models adequately
approximate the NEI models.

In our CIE tests (Section~\ref{subsect:cievsnei}), shock heating
analysis
(Section~\ref{subsect:shockheating}), and 
turbulent mixing analysis (Sections~\ref{subsect:hotism}),
we use a 3 dimensional Cartesian coordinate system with
width (span in the $x$ direction),
and depth (span in the $y$ direction)
of 1.5~kpc and 1.5~kpc, 
respectively.
The footprint is centered on $x = 0$~kpc, $y = 0$~kpc.
Most of these simulations use a domain that is 6~kpc tall.
The exceptions are the high resolution, radiatively cooling, 
shock scenario simulations,
for which we reduced the domain height to 1.5~kpc.
In the turbulent mixing simulations,
hydrostatic equilibrium is achieved by decreasing the thermal 
pressure with height above the midplane.   We set the physical
conditions at the base of the domain to simulate those in the ISM
at $z = 8$~kpc, thus in these simulations
the domain runs from $z = 8$ to 14~kpc.
In the remaining simulations, 
there are no gravitational, density, pressure, or magnetic
field gradients in the modeled ambient gas.
Thus, an arbitrary $z$ offset can be added to the height along
the $z$ axis.
For these simulations, we choose to call the lower boundary 
$z$ = 0~kpc, only for convenience.
In the shock models that had no radiative cooling,
each domain is initially segmented into 4 blocks.
Each block is subdivided by the AMR routine an additional 
3 to 5 times,
with the number of subdivisions depending upon the density 
gradient.   With 5 levels of refinement, the resulting blocks
are 
$1500/(2^{5-1})$~pc,
on a side,
while with 3 levels of refinement they are $1500/(2^{3-1})$ on a side. 
Each of these blocks is segmented into 8$^3$ zones, which, 
consequently, range in size from 
(12~pc)$^3$ 
to (47~pc)$^3$.   
Like the shock models that had no cooling, 
some of our shock models that had cooling also had
moderate resolution and 6~kpc tall domains initially made
from 4 blocks,
while others used higher resolution and
a single 1.5~kpc tall block that was subdivided an additional
3 to 7 times depending upon the density gradient.
These blocks were then subdivided into 8$^3$ zones
that range in size from
(2.9~pc)$^3$ to (47~pc)$^3$.   
For our resolution tests (Section~\ref{subsect:resolution}),
we use both 3 dimensional Cartesian domains and
2 dimensional cylindrically symmetric domains
and test maximum refinement levels between 4 and 8.

We calculate each zone's emission spectrum
using the Raymond and Smith
spectral code (\citealt{raymond_smith_77}, with updates),
the gas density calculated by FLASH, the ionization levels from
either the CIE or the NEI calculations, and
\citet{allen_73} cosmic abundances.
The Allen abundances are greater than the gas phase abundances
in the halo and clouds.
For example, the gas phase metal abundances in the warm halo
are $\sim 4/10$ of the Allen solar abundances
\citep{savage_sembach_96},
while the gas phase oxygen abundance in Complex C is about
$\sim1/10$ to $\sim1/4$ of the Allen solar values \citep{fox_etal_04},
and the gas phase metal abundances in Magellanic Stream clouds
are $\sim 1/14$ of the Allen solar values \citep{fox_etal_10}.
The strength of the emitted spectra should be scaled accordingly,
bearing in mind that in the shock scenario
(the most X-ray productive scenario), 
the X-rays originate in the ambient gas and that over time,
collisions in hot gas break down dust.

In order to obtain the spectrum pertaining to any given 
line-of-sight, we
sum the spectra from the intersected zones.   If the zones
have differing sizes, we weight the terms accordingly.
The X-ray spectra are convolved with the \rosat\ response
matrix in order to determine the count rate in the \rosat\ R12 band,
the 1/4~keV energy band.
For the brighter models, we also report 
intensities at the photon energies of the \oxyseven\ 
triplet ($\sim570$~eV) and \oxyeight\ Lyman alpha line (653~eV).
During our line intensity calculations,
we do not subtract the continuum or the pseudo-continuum 
composed of faint, unresolved lines, because they account for
$\sim5\%$ of the emission in these energy bins during the
brightest phases.   
They account for a greater fraction of
the intensity when the gas is dimmer, but at these times, the
reported intensity is too dim to be observed.
In addition, for comparison with the observed X-ray excess
associated with a Magellanic Stream cloud
(\citealt{bregman_etal_09}; $0.64 \pm 0.10$ counts~ks$^{-1}$~arcmin$^{-2}$
seen by the \xmmnewton\ pn detector in 0.4 to 1.0~keV X-rays),
we calculate the XMM pn count rate for four of our
shock models.

We calculate the density of \oxyseven\ 
ions for select models by combining the fraction of oxygen in the
\oxyseven\ ionization state
with the oxygen abundance 
and the gas density in each zone.
We integrate the density along
sight lines through the domain in order to obtain 
\oxyseven\ 
column densities.    Again, the results can be scaled if
elemental abundances other than those of \citet{allen_73} 
are preferred.
\\

\subsection{Cloud and ISM Parameters}
\label{subsect:parameters}

We initialize the 6 kpc tall domains 
such that a spherical cloud is located
in the upper portion of the grid
and is surrounded by stationary ambient gas.   In the
1.5 kpc tall domains, the center of the cloud is initially
located at 2/3 of the height of the domain.
At the beginning of each simulation, the cloud has a
radius of 0.2~kpc, which corresponds to 1.1\degr\ if the
cloud is seen from 
the current distance to Complex C (10 kpc;  \citealt{thom_etal_08}).
In most of our simulations,
the cloud's initial volume density of hydrogen nuclei,
$n_{cl,H}$, is $0.0645$~cm$^{-3}$.   
Note that the simulations
assume a 10 to 1 ratio of hydrogen to helium, 
making the mass density 
$0.0903$~amu~cm$^{-3}$ and note that
the number density shown in the figures and quoted in
the text is the number of hydrogens per unit volume.
We choose an initial hydrogen volume density of $0.0645$~cm$^{-3}$
because clouds with this density and 
a radius of 0.2~kpc have 
hydrogen column densities along sight lines through 
their centers of
$8 \times 10^{19}$~H~cm$^{-2}$, which
is similar to the observed \hone\ column density in
small features within Complex C.
Initial cloud temperatures range from 
100~K to $\sim11,000$~K and are chosen from the constraint that
the cloud's initial thermal pressure must balance the 
initial thermal pressure of the gas around it.   
In our cases with magnetic fields, the initial magnetic field
pressures also balance.

Table~\ref{table:parameters} lists 
the cloud and interstellar medium parameters for our simulations.
They are grouped into three
general cases, case \hot\ (clouds falling under the influence of
gravity through a hot ambient medium 
that is in hydrostatic equilibrium; all of these test turbulent mixing, 
and some also test shock heating),
case \purehd\ (fast clouds moving through hot, rarefied gas before
colliding with warm, moderate density ambient gas; these test shock
heating and there are two categories, those with radiative cooling (\nei)
and those without (\adiabatic)),
and case \warmism\ (fast clouds moving through warm, moderate density
ambient gas; these test shock heating).

\begin{deluxetable*}{ll|cccccc}
\tabletypesize{\small}
\tablewidth{0pt}
\tablecaption{Simulation Parameters}
\tablehead{
\colhead{Model}
& \colhead{Comment}
& \colhead{$n_{cl}$}
& \colhead{$T_{cl}$}
& \colhead{$n_{ISM}$} 
& \colhead{$T_{ISM}$} 
& \colhead{$v_z$} 
& \colhead{$B_y$}
\\
\colhead{  }
& \colhead{  }     
& \colhead{(H cm$^{-3}$)}
& \colhead{(K)}
& \colhead{(H cm$^{-3}$)} 
& \colhead{(K)} 
& \colhead{(km~s$^{-1}$)} 
& \colhead{($\mu$G) } 
}
\startdata
\hot 1  & note a & $6.45 \times 10^{-2}$ & $7.83 \times 10^{2}$ & see Figure~\ref{fig:density}:   & $10^6$  & $0$ & 0  \\
\hot 2  & '' $h = 11$~kpc & '' & '' & '' & ''  & $-50$ & ''  \\
\hot 3  & note a & $6.45 \times 10^{-3}$ & $7.83 \times 10^{3}$ & '' & ''  & $0$ & ''  \\
\hot 4  & '' & $6.45 \times 10^{-2}$ & $7.54 \times 10^{3}$ & '' & $3 \times 10^6$  & '' & ''  \\
\hot 5  & '' & '' & $1.11 \times 10^{3}$ & '' & $10^6$  & '' & '' \\
\hot 6  & '' & '' & $7.83 \times 10^{2}$ & '' & ''  & '' & 0.1  \\
\hot 7  & '' & '' & '' & '' & ''  & '' & 0.5  \\
\hot 8  & '' & '' & $1.11 \times 10^{3}$ & '' & ''  & $-300$ & 0 \\  
\hot 9  & '' & '' & '' & '' & ''  & $-400$ & '' \\  
\hot 10 & '' & $6.45 \times 10^{-3}$ & $1.11 \times 10^{4}$ & ''& ''  & $-300$ & '' \\  
%
\hot 11& '' & $0.15$ & $4.6 \times 10^3$  & ''& $3 \times 10^6$ & 0 & '' \\  
\adiabatic 2 & notes b and ba & $6.45 \times 10^{-2}$ & $10^3$	& $6.45 \times 10^{-3}$ & $10^4$ & $-200$	& '' \\
\adiabatic 3 & '' & '' & '' & '' & '' & $-300$	& '' \\
\adiabatic 4 & '' & '' & '' & '' & '' & $-400$ & '' \\
\adiabatic 5 & '' & '' & '' & '' & '' & $-500$	& '' \\
\adiabatic 6 & '' & '' & '' & '' & '' & $-600$	& '' \\
\nei 2 &  notes b and br & '' & ''	& '' & '' & $-200$	& '' \\
%
\nei 3   &  '' & '' & ''	& '' & '' & $-300$	& '' \\
%
\nei 4   &  '' & '' & '' 	& '' & '' & $-400$	& '' \\
\nei 3d  &  '' & '' & $10^2$	& $6.45 \times 10^{-4}$ & '' & $-300$	& '' \\
%
\warmism 3  & note c & '' & $10^3$ & $6.45 \times 10^{-3}$ & ''  & $-300$ & '' \\
\warmism 4  & '' & '' & '' & '' & ''  & $-400$ & '' \\
\warmism 5  & '' & '' & '' & '' & ''  & $-500$ & '' \\
\warmism 6  & '' & '' & '' & '' & ''  & $-600$ & '' \\
\warmism 3d  & '' & '' & $10^2$ & $6.45 \times 10^{-4}$ & '' & $-300$ & '' \\
\warmism 6d  & '' & '' & '' & '' & ''  & $-600$ & '' \\
%
\enddata
\tablecomments{ \\
note a:  Unless otherwise stated, the clouds in the suite
\hot\ models begin at height $h = 12$~kpc.
The temperature at the cloud's initial center is listed, but
the cloud's 
temperature varies slightly with height in order to balance the cloud's 
pressure with that of the ambient gas.
In models \hot1 through \hot11, the thermal pressures at the 
midplane 
are 6000, 6000, 6000, 4190, 8480, 6000, 6000, 8480, 8480, 8480,
and 5930~K~cm$^{-3}$, respectively.   
All suite \hot\ simulations
assume CIE ionization levels and no radiative cooling.   
\\
note b:  The tabulated temperature and density refer to those 
of the warm ISM; the hot ISM, occupying the upper 1.5~kpc (in 
the Model \adiabatic\ simulations) 
or 1.0~kpc (in the Model \nei\ simulations) of the 
domain has a temperature of $10^6$~K and a density of 1/100 of 
that in the warm ISM. \\
note ba:  All suite \adiabatic\ simulations assume CIE ionization
levels and no radiative cooling. \\
note br:  All suite \nei\ simulations modeled NEI ionization
fractions for select elements during their FLASH runs and CIE
ionization fractions in the post processing.   In addition
all \nei\ simulations include radiative cooling. 
We made both moderate and high resolution versions of the 
Model \nei\ simulations.\\
note c:
All suite \warmism\ simulations
assume CIE ionization levels and no radiative cooling.   
}
\label{table:parameters}
\end{deluxetable*}

\begin{figure}
\epsscale{1.2}
\plotone{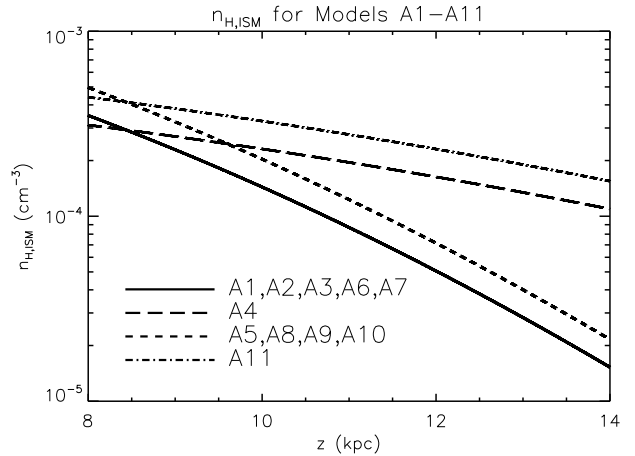}
  \caption{
Density of ambient gas, expressed in units of hydrogen
atoms per cm$^{-3}$, as a function of height
above the Galactic plane for the models that have hydrostatic 
environments, i.e., Models \hot 1 through \hot 11.
}
\label{fig:density}
\end{figure}

In our suite of case \hot\ simulations,
the ambient medium is 
either $T_{ISM} = 1 \times 10^6$ or $3 \times 10^6$~K and,
as mentioned in Subsection~\ref{subsect:mhd}, radiative cooling
is disabled in order to maintain hydrostatic equilibrium in the
ambient gas.
We determine the initial density gradient
from the equation for hydrostatic equilibrium 
with the constraints that the magnetic field
strength and temperature are constant.
We constrain the midplane density by requiring the 
midplane thermal pressure to be reasonable, i.e. roughly
6000~K~cm$^{-3}$, and the 
1/4~keV X-ray surface
brightness of the ambient medium above $z = 2$~kpc to be
comparable to observationally derived values.
As an example, we note the X-ray surface brightness
in the 1/4~keV \rosat\ band 
produced by the gas above 2~kpc 
in our Model~\hot\ reference run, \hot1.
Its intrinsic surface brightness 
is $1510 \times 10^{-6}$ counts s$^{-1}$ arcmin$^{-2}$.
If the
column density of intervening \hone\ is $1 \times 10^{20}$~cm$^{-2}$,
then the absorbed surface brightness is 
$583 \times 10^{-6}$ counts s$^{-1}$ arcmin$^{-2}$,
which is consistent with measurements 
presented in
\citet{snowden_etal_98,snowden_etal_00}.
Our constraints yield the ambient density functions plotted
in Figure~\ref{fig:density}; thus the density of ambient gas,
expressed in units of hydrogens per unit volume, at the 
cloud's initial height in most of our Case \hot\ simulations
is between $\sim5 \times 10^{-5}$ and $2.5 \times 10^{-4}$~cm$^{-3}$.
The values are consistent with those used or determined
elsewhere (e.g., density of $\ga 2.4 \times 10^{-5}$~cm$^{-3}$,
(\citealt{blitz_robishaw_00}, ram pressure stripping of nearby
dwarf galaxies);
$\sim5 \times 10^{-5}$~cm$^{-3}$, (\citealt{moore_davis_94},
ram pressure stripping of Magellanic Clouds);
$2 \times 10^{-4}$~cm$^{-3}$ 
in model by \citet{blandhawthorn_etal_07};  1 and
$3 \times 10^{-4}$~cm$^{-3}$ in models by \citet{heitsch_putman_09}). 
In order to balance the ambient and cloud pressures at the beginning
of each simulation, we allow the cloud temperature to vary slightly
with height.
We make a suite of simulations in order to sample various 
initial cloud heights and velocities,
ambient densities and temperatures, and
strengths of the
magnetic field lying parallel to the Galactic midplane.
All of the Case \hot\ models use CIE calculations in order
to determine the fraction of atoms in any given ionization
state.
The hydrodynamical results from Model \hot\ are stored at 2~Myr
intervals.

In our suite of Case \purehd\ simulations, 
the clouds pass through a hot ambient medium before 
hitting a warm extended medium.  The hot medium may represent halo
or intergalactic gas, while the warm medium may represent 
material shed from a preceding HVC, a high $z$ cloud, or, 
given the proximity of some HVCs to the Galactic disk,
it may represent warm gas in the Galaxy's thick disk.   
In our set-up, the hot
and warm sectors are in pressure balance with each other.
In most of our Model \purehd\ simulations, the hot medium has a 
temperature of $T_{ISM} = 1 \times 10^6$~K,
density of hydrogen of 
$n_{ISM,H} = 6.45 \times 10^{-5}$~cm$^{-3}$, i.e.
$n_{cl,H}/1000$, and occupies the uppermost 1.5~kpc of the grid.
The density of the hot portion of the domain is equal to that 
at $z \sim 12$~kpc in Models \hot 1, 2, 3, 6, and 7.
In most of our Model \purehd\ simulations, the warm part of the 
domain has a temperature of $T_{ISM} = 10^4$~K,
density of hydrogen of 
$n_{ISM,H} = 6.45 \times 10^{-3}$~cm$^{-3}$, i.e.
$n_{cl,H}/10$, and occupies the lower 4.5~kpc of the grid.
At the beginning of each Case B simulation, 
the environmental gas is in pressure balance with the cloud.
There is no magnetic field and therefore no magnetic pressure.
It is not practical to model hydrostatic equilibrium in 
Case \purehd\
because the weight of the warm medium cannot be supported
unless the pressure
has a strong gradient.
By giving up hydrostatic equilibrium,
we also give up gravity.   
Because both are absent,
an arbitrary offset can be added to our domain's $z$ without
affecting the results.  
On this arbitrary scale,
the initial location of the center of the 
cloud is $z = 5$~kpc in simulations
having 6~kpc tall domains (the moderate resolution simulations)
and 0.5~kpc from the top in the simulations having 1.5~kpc tall domains
(the high resolution simulations).
Lacking gravity to accelerate them, 
the clouds must be given an initial velocity.   We 
sample several velocities; 
see Table~\ref{table:parameters}.  
Cases with speeds of 
200 and 300~km~s$^{-1}$ can be used for some Galactic HVCs.
Cases with speeds of
300 and 400~km~s$^{-1}$ can be compared with
the Magellanic Stream, whose orbital speed is 
$378 \pm 18$~km~s$^{-1}$ \citep{blandhawthorn_etal_07}.
The simulations with faster clouds are useful for identifying
trends.
We test the effects of non-equilibrium ionization(NEI) by
creating a small suite of models 
and for which both NEI and CIE ionization fractions are
calculated  (these models have names that begin with \nei\
and include radiative cooling).   We test the effects of 
radiative cooling at the CIE rate by comparing models with
and without radiative cooling.   Those without radiative
cooling have names that begin with \adiabatic.
The computational results for Case \adiabatic\ models
are stored at 2~Myr intervals.  
We performed several Model \nei\ simulations with the same 
resolution and archival time intervals as the Model \adiabatic\ 
simulations, but in order to better model the radiative
cooling in this scenario, simulations having smaller 
zone sizes and stored time intervals are also needed.
Thus, we also present simulations of \nei\ models
in which up to 
7 levels of refinement are allowed, the spacing between 
archived epochs is 40,000~years, and the simulations are
stopped at 2~Myr.   It is this set of simulations that has
1.5~kpc domain heights.

Case \warmism\ is a variation on Case \purehd, with the primary 
difference
being that all of the ambient gas is warm ($T_{ISM} = 1 \times 10^4$~K) 
in Case \warmism.  
In Case C, we also sample two ambient densities,
a rarefied case having 
$n_{ISM,H} = 6.45 \times 10^{-4}$ H cm$^{-3}$ and
a moderate case having the same density as the warm
gas in Case \purehd, 
i.e. $n_{ISM,H} = 6.45 \times 10^{-3}$ H cm$^{-3}$.
In the former case, we also reduce the clouds'
temperature in order to achieve pressure balance 
at the beginning of the simulations.   
All of the Case C models use CIE calculations, 
omit radiative cooling, and are archived every 2~Myr.
\\

\section{Results}
\label{sect:results}

\subsection{Validity of CIE Approximation}
\label{subsect:cievsnei}

\begin{figure}
\epsscale{1.2}
\plotone{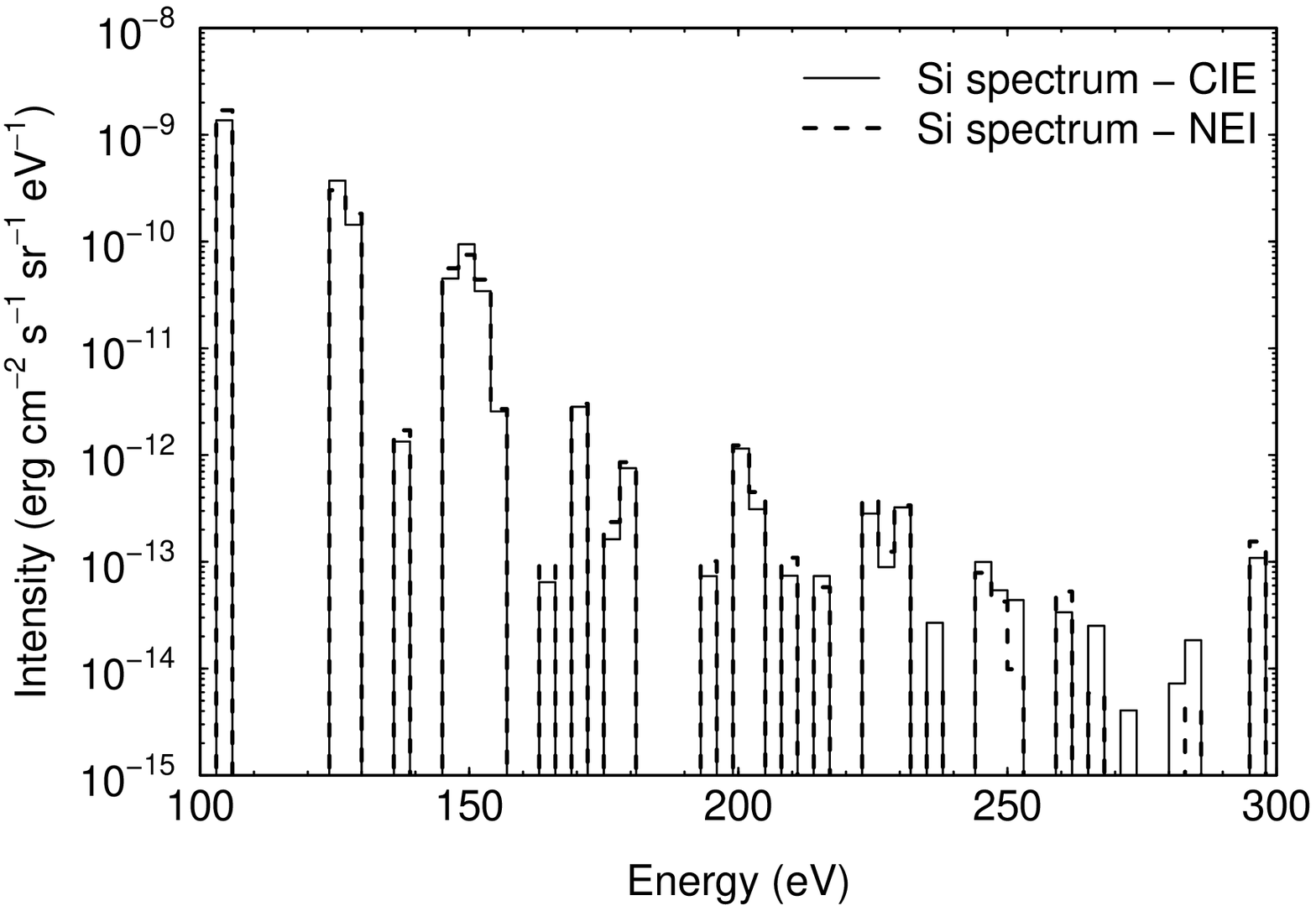}
\plotone{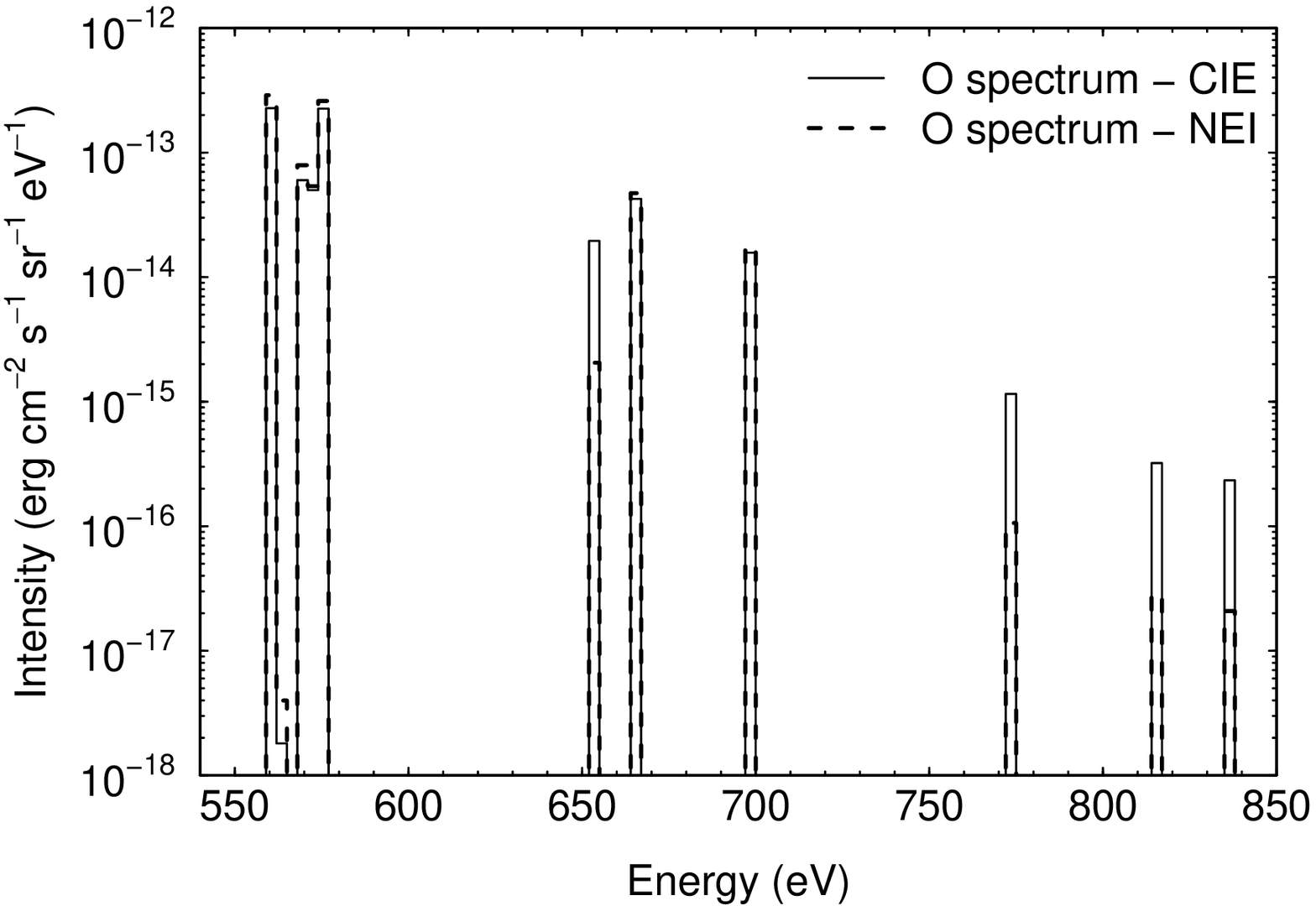}
  \caption{ 
Top:  The emission line spectra of silicon, calculated using the 
NEI algorithm and
the CIE algorithm, produced along the central sight line in the 
moderate resolution \nei 4 model 
at 2~Myr age.   
Bottom: The NEI and CIE emission line spectra produced by oxygen ions 
along the same sight line.
This plot shows the similarity between the CIE and NEI spectra 
from silicon and oxygen, important contributers to
the 1/4 keV and 3/4 keV count rates, respectively.}
\label{fig:cievsnei}
\end{figure}

In order to evaluate the CIE approximation, we compare spectra
that we calculate from the NEI and CIE ionization fractions
of silicon and oxygen.
The NEI ionization fractions, i.e., the fractions of the atoms 
at any given ionization level
were obtained by having FLASH track
the ionization levels in a time dependent fashion in moderate
resolution versions of models
\nei 3, \nei 4, and \nei 5.
The CIE ionization fractions were obtained from the hydrodynamic
information for these same models;  specifically, they were 
calculated by the Raymond and Smith code using the 
temperature of the gas as an input.
For each of these three models, we calculate the spectra
produced by NEI silicon, CIE silicon, NEI oxygen, and CIE oxygen
for various vertical sight lines through the domain at each epoch,
2~Myr, 4~Myr, etc. (Although the shocked gas has already
radiated way much of its energy by 2~Myr, it is still
sufficiently emissive for these experiments.)
In order to reduce the quantity of information, we
then convolve each silicon spectrum with the 
\rosat\ 1/4 keV band response function 
and each oxygen spectrum with the 
\rosat\ 3/4 keV band response function, 
yielding count rates in these bands.
In each case, 
the CIE count rates in the 1/4 keV band
are fairly similar to the NEI count rates.
For example, the \rosat\ 1/4 keV count rate calculated from the 
CIE and NEI silicon spectra 
produced along the central sight line of the \nei 4
model at 2~Myr are 
within $5\%$ of each other.
The relationship between the CIE and NEI oxygen spectra were
not as consistent from model to model and epoch to epoch. 
For Model \nei 4 at 2~Myr the CIE and NEI oxygen
spectra are within $14\%$ of each other.
The emission line spectra for this case
are displayed in Figure~\ref{fig:cievsnei}.   
Sometimes, the \oxyseven\ intensities from the models
are less consistently aligned with each other
than are the silicon intensities and frequently the
\oxyeight\ quantities (which are always small)
are poorly aligned, presumably due to
delayed ionization and recombination.  Similarly, the high
resolution Model \nei\ simulations also show poor alignment
between the CIE and NEI \oxyeight\ column densities.\\

\subsection{Shock Heating}
\label{subsect:shockheating}

\begin{figure*}
\centering
\includegraphics[scale=0.3]{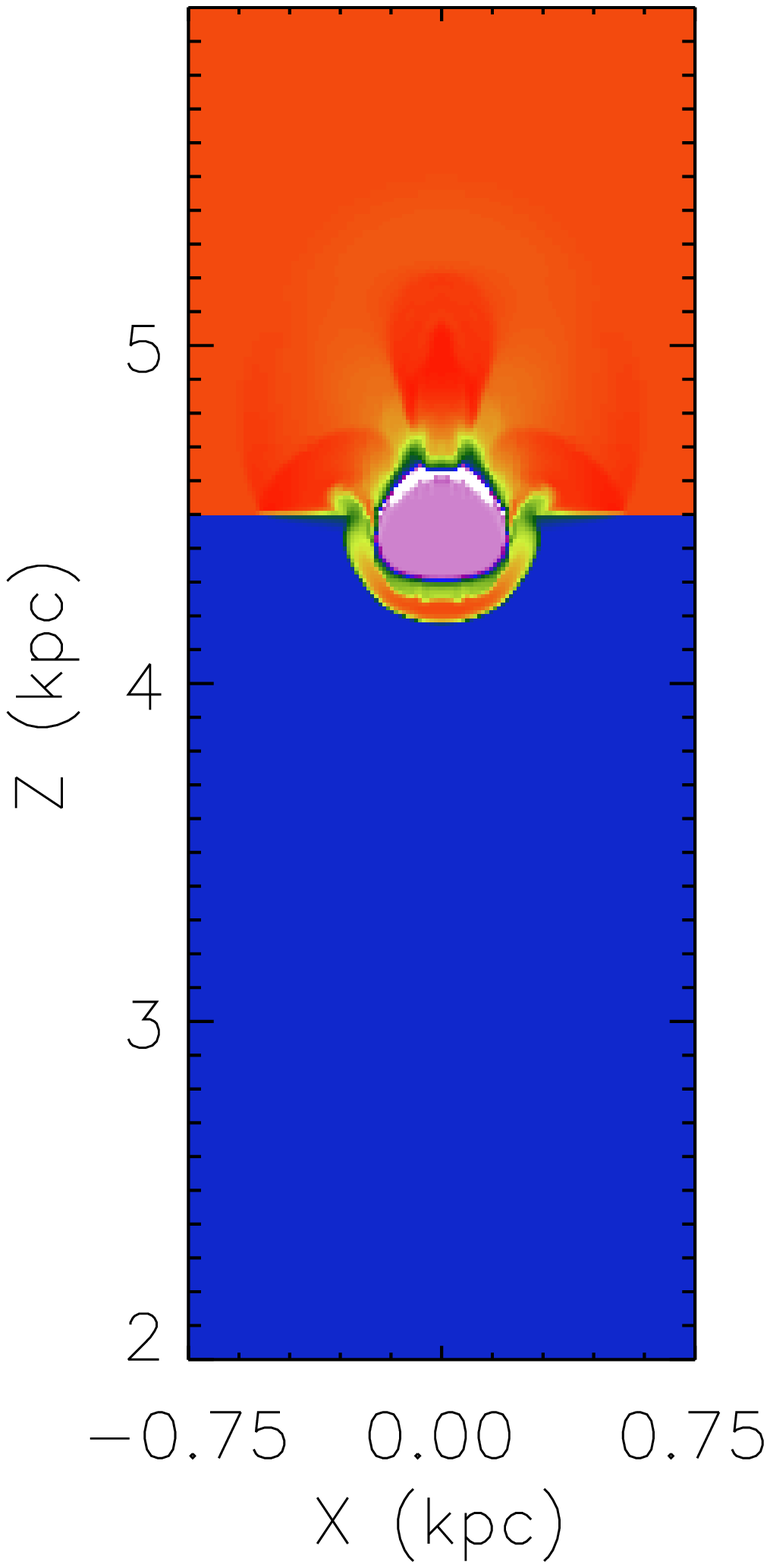}
\hspace{0.25in}
\includegraphics[scale=0.3]{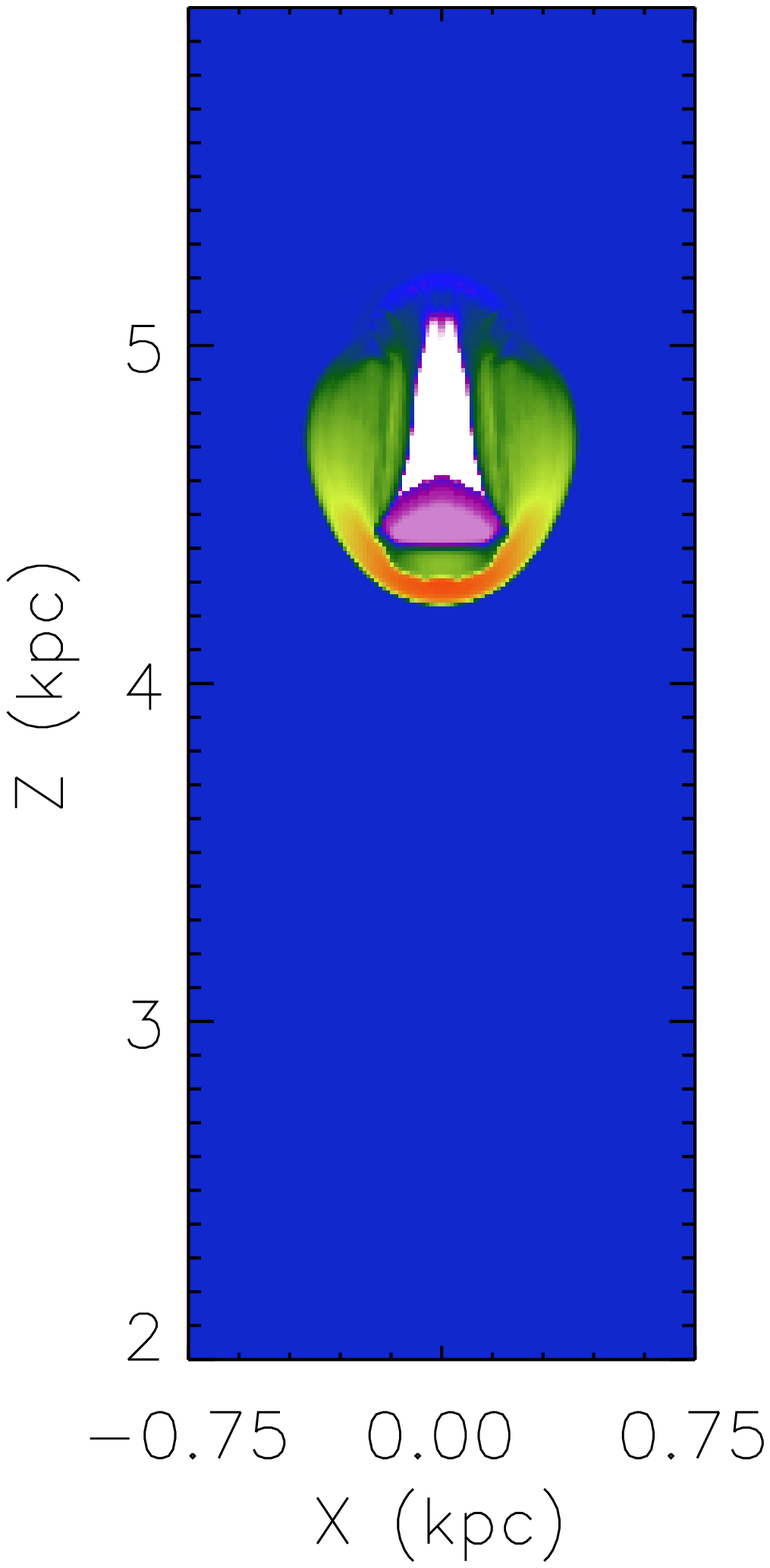}
\hspace{0.25in}
\includegraphics[scale=0.3]{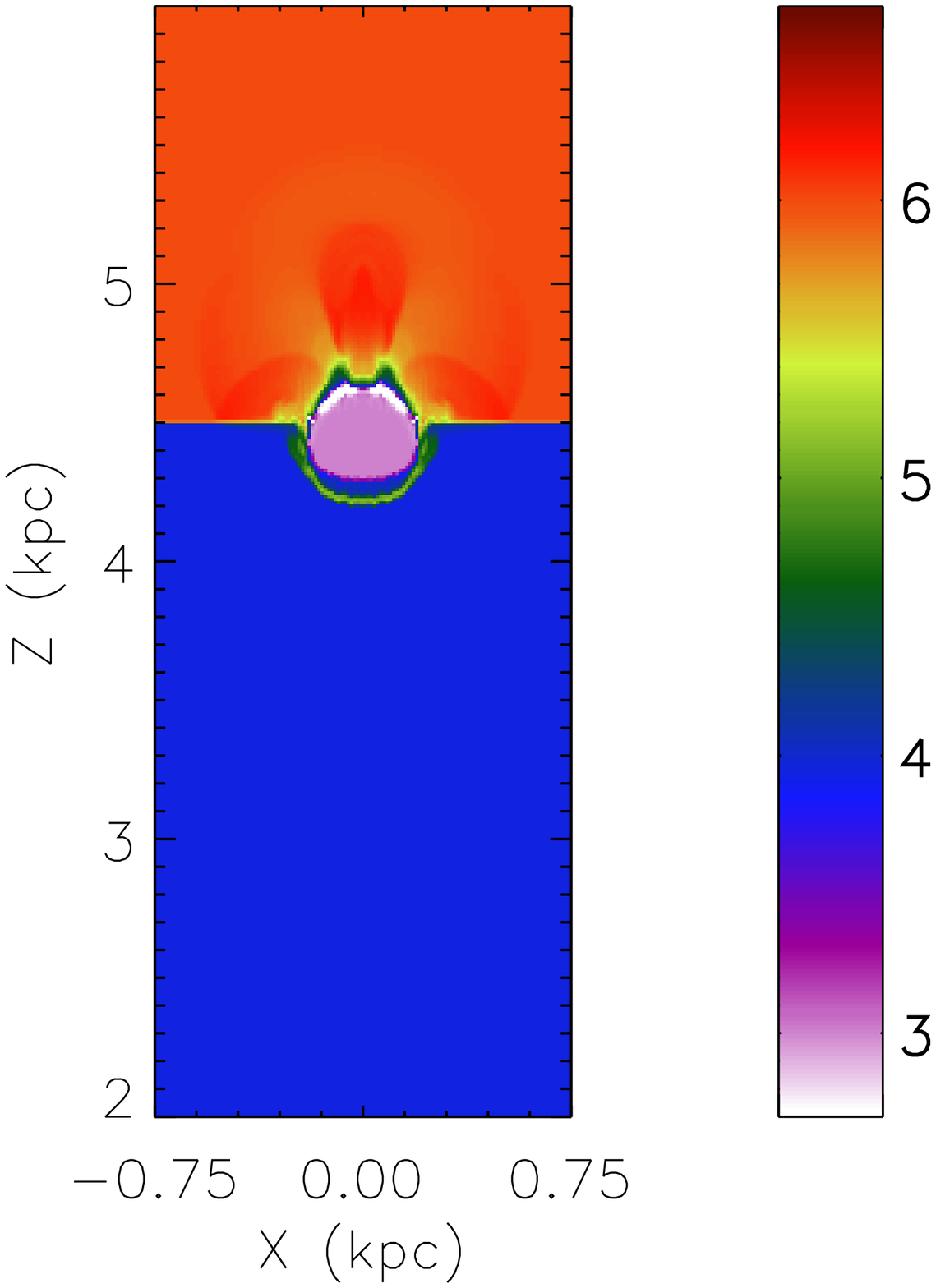} \\
\vspace{0.5in}
\includegraphics[scale=0.3]{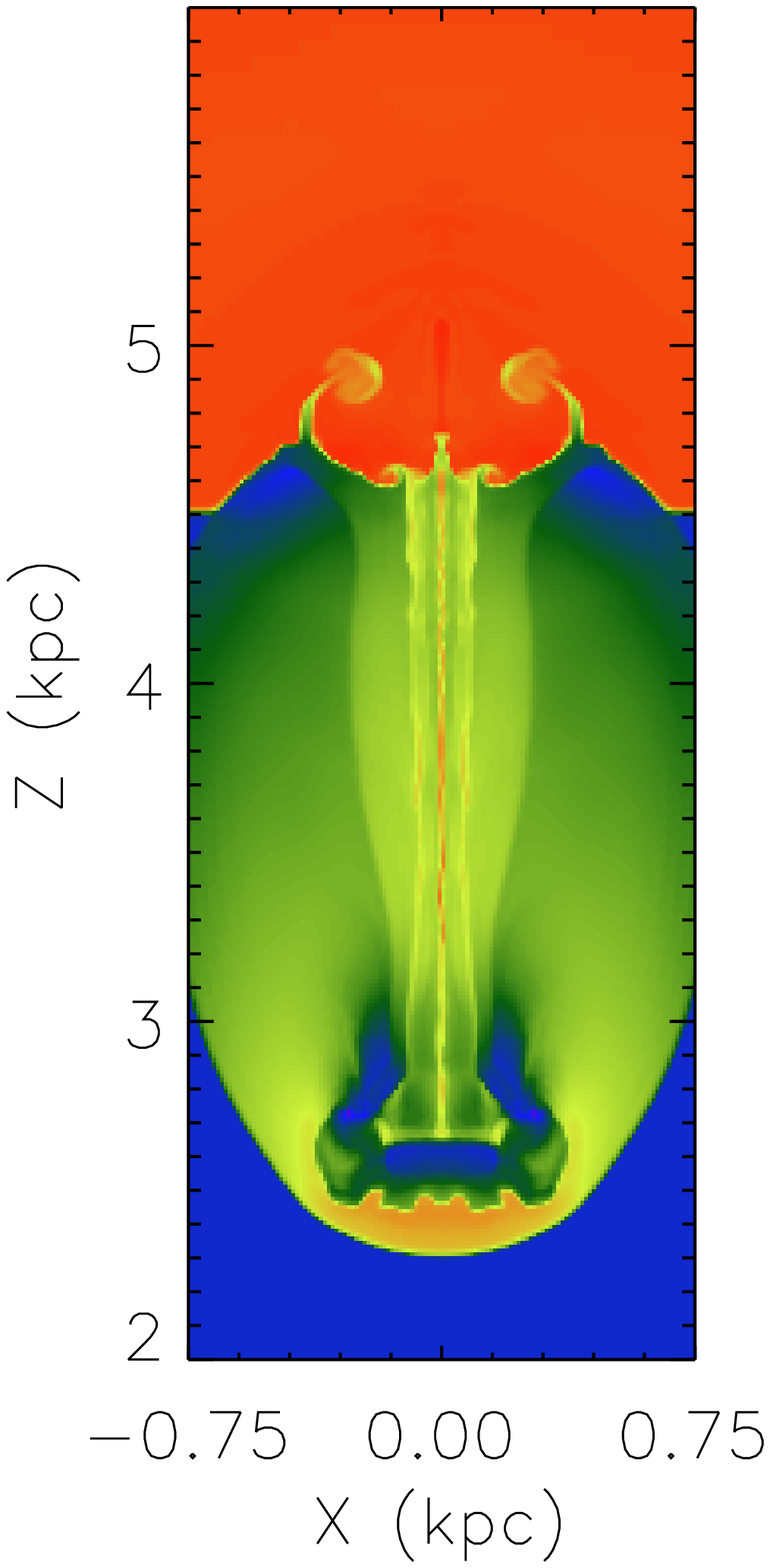}
\hspace{0.25in}
\includegraphics[scale=0.3]{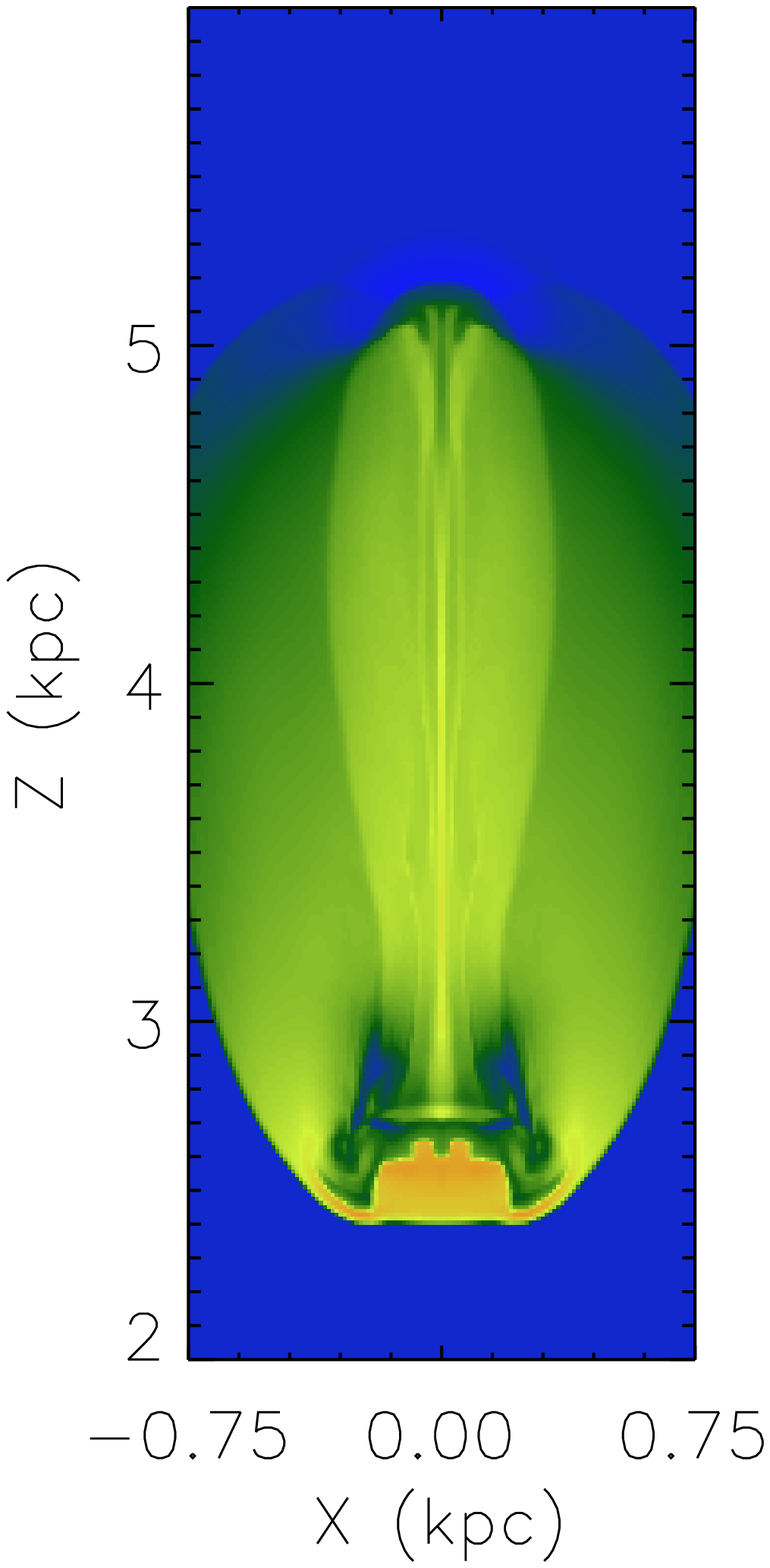}
\hspace{0.25in}
\includegraphics[scale=0.3]{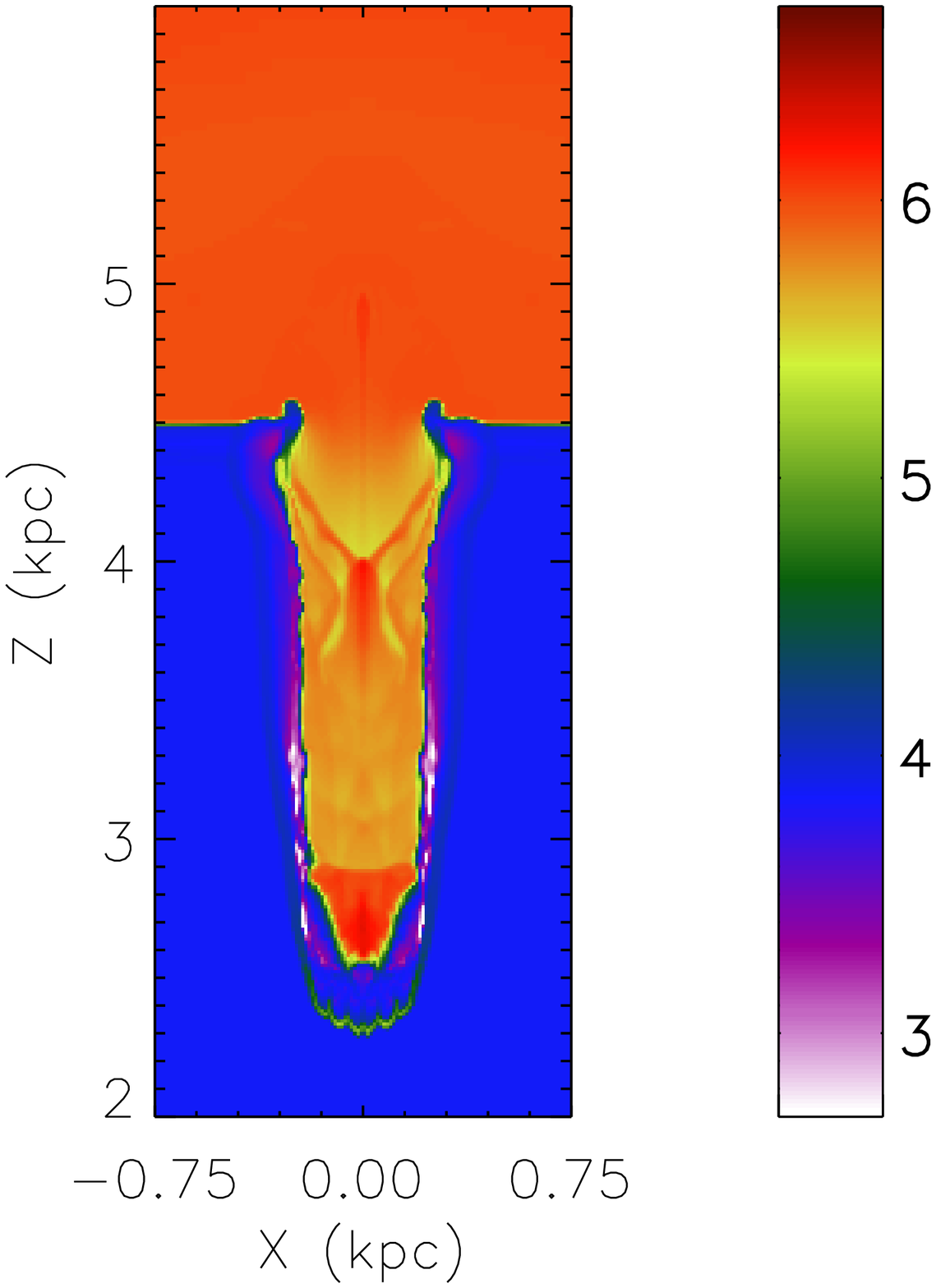}
\vspace{0.5in}
  \caption{   
The log of the temperatures of the cloud and ambient gas along a 
vertical slice through the centers of models \adiabatic 3,
\warmism 3, and \nei 3 (left to right)
at 2~Myr (top row) and at
10~Myr (bottom row).
The color bars are keyed to the log of the temperature.
The cloud appears at $z \sim 4.5$~kpc in the top images and at 
$z = 2.6$~kpc in the bottom images.   
Only the upper 4~kpc of the domains are shown.    
The temperature elevations that resulted from the forward shock
in the ambient gas and the reverse shock in the cloud can
be seen in the \adiabatic 3 and \warmism 3 images.    
In contrast, the 
\nei 3
image shows little hot gas beneath the cloud, owing to
radiative energy losses.  In order to show the late-time 
evolution for Model \nei3, we used our moderate resolution
simulations.
The apparent pixelation in the images
is an artifact of the image processing 
and does not represent the resolution of the simulations.
\\
}
\label{fig:images}
\end{figure*}

Theoretically, if the clouds collide with neighboring gas
at fast enough speeds, then shocks should develop in both
media.
The temperature ($T_2$) of the shocked plasma 
can be estimated from the equation
for plane parallel shocks \citep{shu_92}:
\begin{equation}
	T_2 = T_1 \frac{[(\gamma+1)+2\gamma(M_1^2-1))]
	[(\gamma+1)+(\gamma-1)(M_1^2-1)]}{(\gamma+1)^2 M_1^2},
       \label{eq:shocktemp}
\end{equation}
where $T_1$ is the temperature of the unshocked gas, 
$M_1$ is the Mach number calculated from the speed
at which the unshocked gas is overrun by the shock,
and $\gamma$ is the adiabatic constant for monotonic
gas ($5/3$).
Here, we compare the theoretical post-shock temperature
with those of Models \adiabatic 3 and \warmism 3
at the first epoch ($t = 2$~Myr).  
These models 
begin with cloud velocities of $v_z = -300$~km~s$^{-1}$ and
do not model radiative cooling (although, in 
post-processing, we can estimate the X-ray spectra that 
would be emitted by gas at the simulated temperatures).
By the first archived epoch at 2~Myr of simulated time, 
the cloud has shock-heated the environmental gas
beneath it while a reverse shock has been driven into
the lower part of the cloud.   In both Model \adiabatic 3
and \warmism 3, the shocked environmental 
gas was initially warm ($T = 1 \times 10^4$~K), ionized ISM.
Although the unshocked portion of the cloud still travels
with approximately its initial velocity, $v_z = -300$~km~s$^{-1}$,
at this epoch, 
the shocked portion of the cloud moves downwards at speeds between
240 and 300~km~s$^{-1}$ and
the swept up ISM moves downwards at speeds up to
240 km s$^{-1}$ with an average of about 200 km s$^{-1}$ 
and moves sideways with speeds exceeding 100 km s$^{-1}$.
Thus, the impacted material does not simply pile up in front of the
cloud, as in a shock-tube simulation, but instead partially 
skirts around the cloud.
The shockfront travels downwards at 4/3 of the shocked ISM's
downwards speed, which is
confirmed by the displacement of the shockfront between $t = 2$ and
4~Myr.
Thus, $M_1 = 18$.   From equation~\ref{eq:shocktemp}, we
find that 
$T_2$ should be $1.0 \times 10^6$~K, which is consistent
with the post-shock temperatures 
in Models \adiabatic 3 and \warmism 3: 0.95 and 
$1.0 \times 10^6$~K, respectively.  See Figure~\ref{fig:images}.
Since the
clouds decelerate in our gravity-less simulations, the
post-shock velocity decreases with time and thus the
post-shock temperature decreases with time as well.
This can be seen by comparing the temperatures at 10~Myr with
those at 2~Myr in Figure~\ref{fig:images}.

A reverse shock propagates through the cloud, but due
to the density contrast, it propagates slower 
than the forward shock in the ISM.
In our simulations of model \warmism 3 at 2~Myr, the reverse shock 
moves into the cloud at a speed of 
$\sim90$ km~s$^{-1}$ from the cloud's reference frame.
Given an initial cloud temperature of $T_{cl} = 1000$~K, 
the reverse shock should be strong and
according to Equation~\ref{eq:shocktemp}, it
should heat the cloud to $\sim1.2 \times 10^5$~K.
Our simulational results find the temperature to range
from the cloud temperature to this value;  see Figure~\ref{fig:images}.
While the reverse shock-heated material is far hotter than the
unshocked portion of the cloud,
it is still not hot enough to produce X-rays.
The X-rays that result from this geometry originate in 
the shock-heated ambient medium.   In the following subsections,
we explain that their intensity ranges from unobservably dim
to significantly bright, depending upon the time since interaction,
collision speed, and strength of radiative cooling.
\\

\subsubsection{Effect of Radiative Cooling}
\label{subsect:cooling}

The simulations and theory discussed above did not include
the effects of radiative cooling.   When it is included in
the calculations, the shocked ambient gas quickly cools.   
This can be seen by comparing the temperature and density 
distributions
in Models \adiabatic 3 and \warmism 3 with those from Model
\nei 3
in Figure~\ref{fig:images}. 
These are fair comparisons because
each model has the same parameters for the cloud and lower ISM 
and, although the ionization fractions of some elements
were calculated in a time-dependent manner in Model \nei 3
(but not in the others), the NEI 
ionization levels did not affect the hydrodynamics and we replace
them with the CIE ionization levels for the following spectral 
calculations.
Furthermore, we note that the ambient medium in 
Models \adiabatic 3 and \nei 3 include a 
layer of hot gas, but this layer contributes only 
	$1 \times 10^{-6}$ counts s$^{-1}$ arcmin$^{-2}$
which is trivially small and is subtracted before we
report the results.

\begin{figure}
\epsscale{1.2}
\plotone{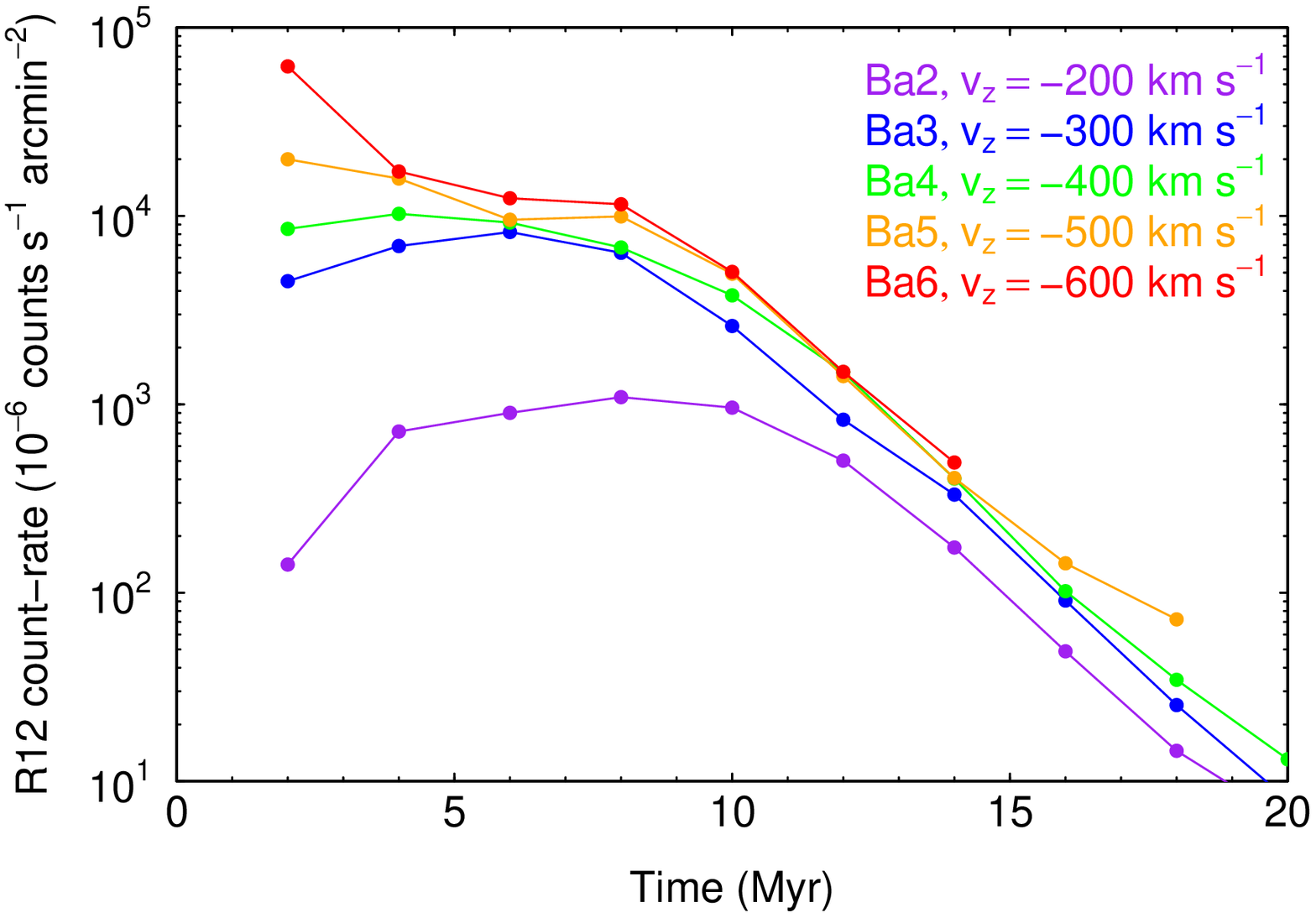} \\
\plotone{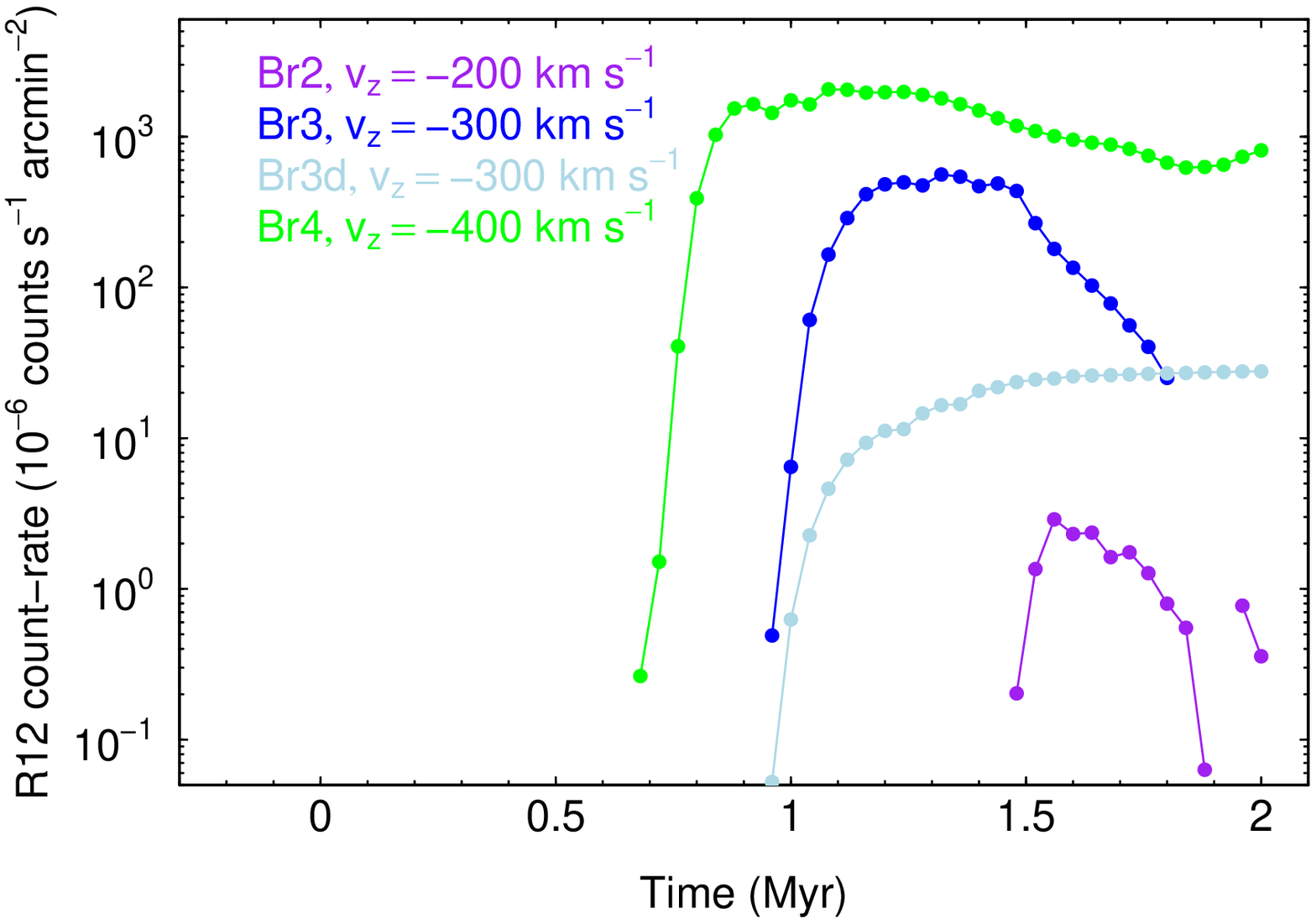} \\
  \caption{
Surface brightnesses in the \rosat\ 1/4 keV band from 
Models \adiabatic 2, \adiabatic 3, \adiabatic 4, \adiabatic 5, 
and \adiabatic 6 and the high resolution versions of 
\nei 2, \nei 3, \nei 3d, and \nei4
as functions of time.   The surface brightnesses are averaged 
across a disk of radius 200~pc and given in units of 
$10^{-6}$ \rosat\ 1/4~keV counts~s$^{-1}$~arcmin$^{-2}$.   
The contribution made by undisturbed hot, background gas
in the domain has been subtracted in order to obtain the
reported results.
The expected intensity due to the background gas 
is brighter than the emission in 
Model \nei 2 at $t = 1.9$~Myr, resulting in unplotted negative 
count rates at that time.
Due to their similarity with the
case \adiabatic\ models, the case \warmism\ 
models are not shown.}
\label{fig:1/4kevevolution} 
\end{figure}

Without radiative cooling, the shock-heated gas is very bright
for millions of years.  As can be seen in 
Figure~\ref{fig:1/4kevevolution}, even the moderately
fast clouds ($v_z = 200$~km~s$^{-1}$) induce
surface brightnesses across their footprints of 
$\sim 1000$ $\times 10^{-6}$ counts s$^{-1}$ arcmin$^{-2}$ in the
\rosat\ R12 band at their peak.
Faster clouds produce 
$\ga 10,000 \times 10^{-6}$ counts s$^{-1}$ arcmin$^{-2}$ in 
this band at their peaks.   These models are much
brighter than the typical intrinsic intensity
of 1/4 keV photons emitted above the Galactic disk, which is
$\sim1000$ to $\sim2000 \times 10^{-6}$ 
counts s$^{-1}$ arcmin$^{-2}$ with significant 
directional variation \citep{snowden_etal_00}.
All of these simulated clouds continue to make
$>100 \times 10^{-6}$ counts s$^{-1}$ arcmin$^{-2}$ 
until at least 14~Myr after the collision.

\begin{figure}
\epsscale{1.2}
\plotone{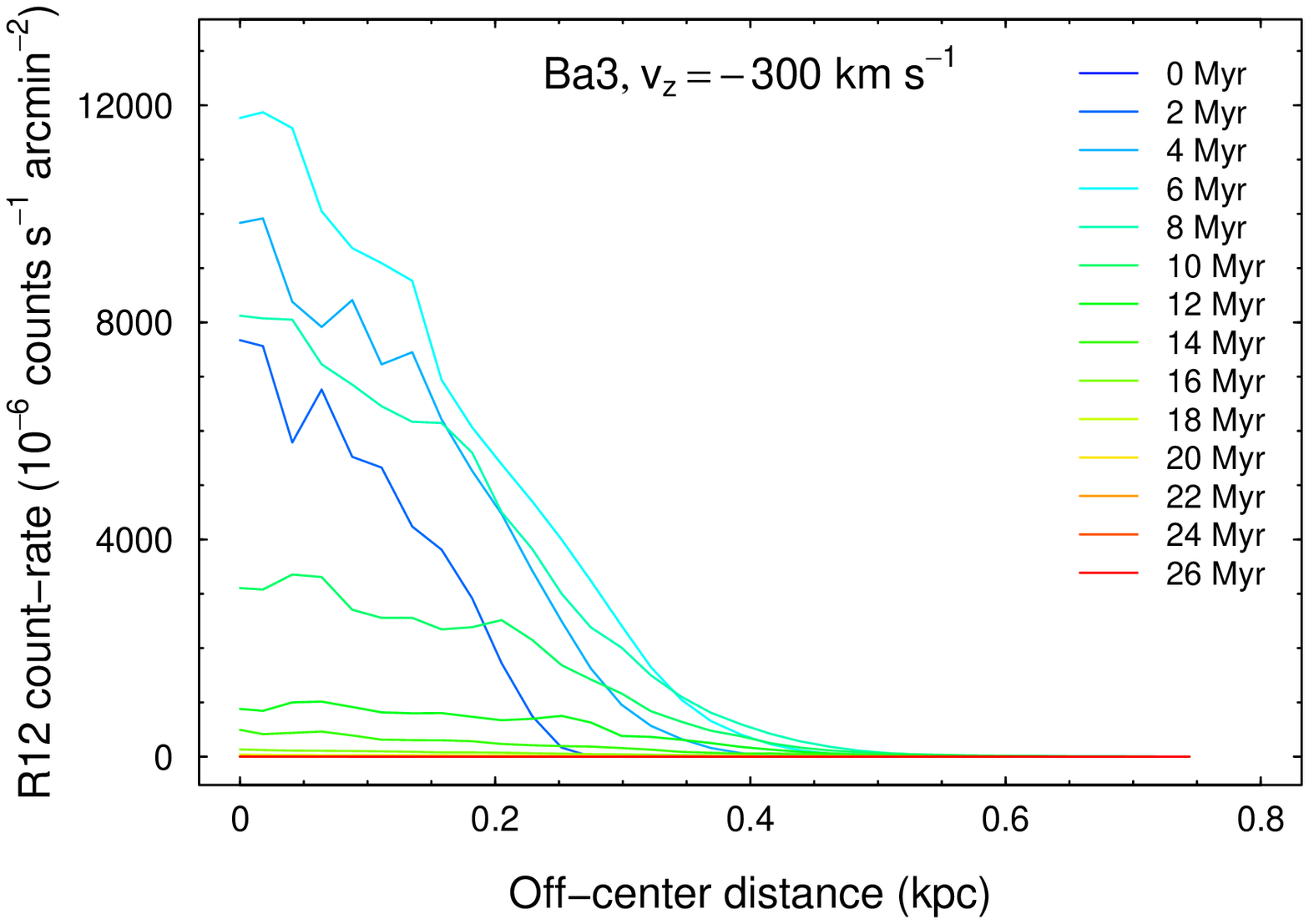}
\vspace{-0.2cm}
\plotone{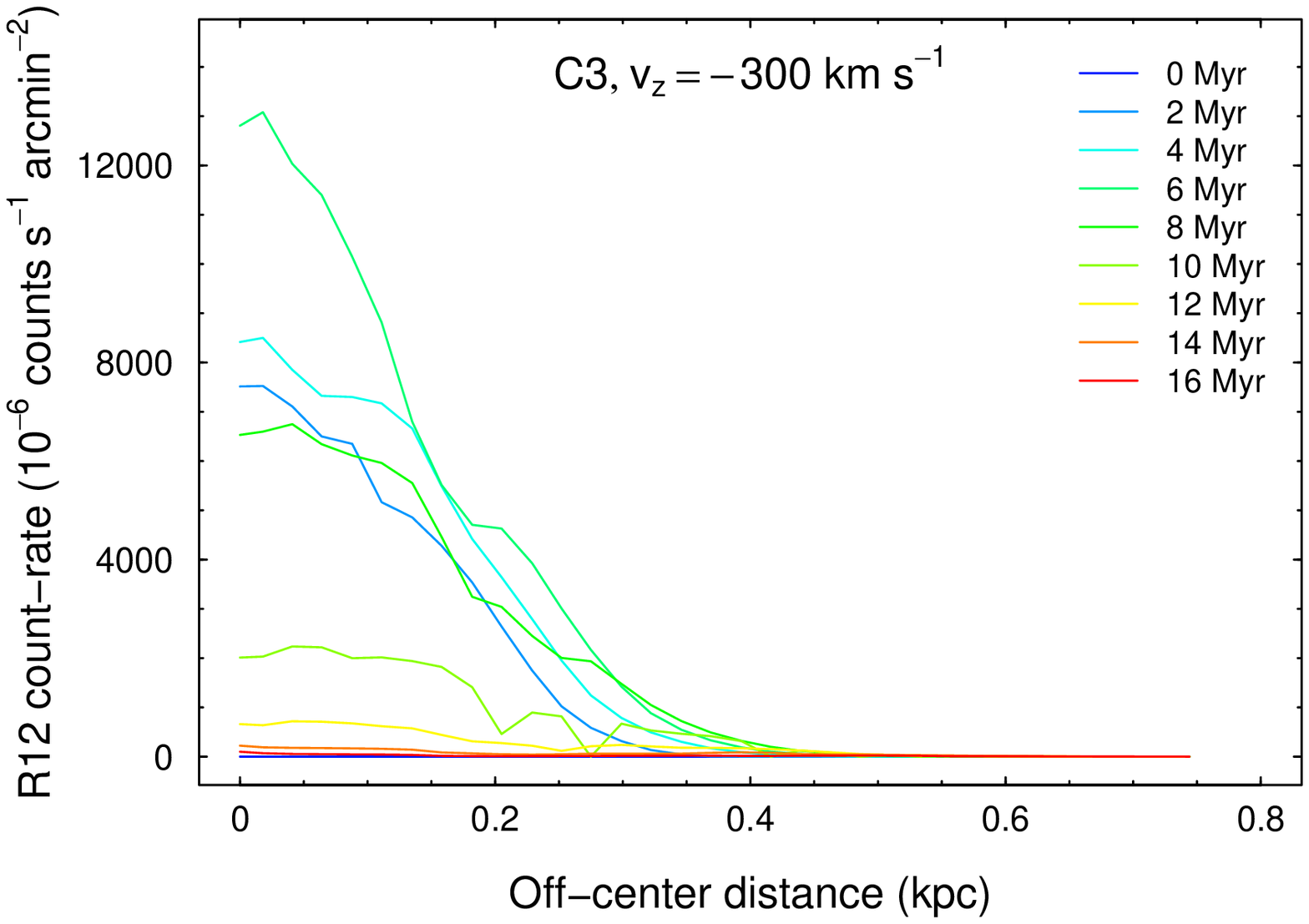}
\vspace{-0.2cm}
\plotone{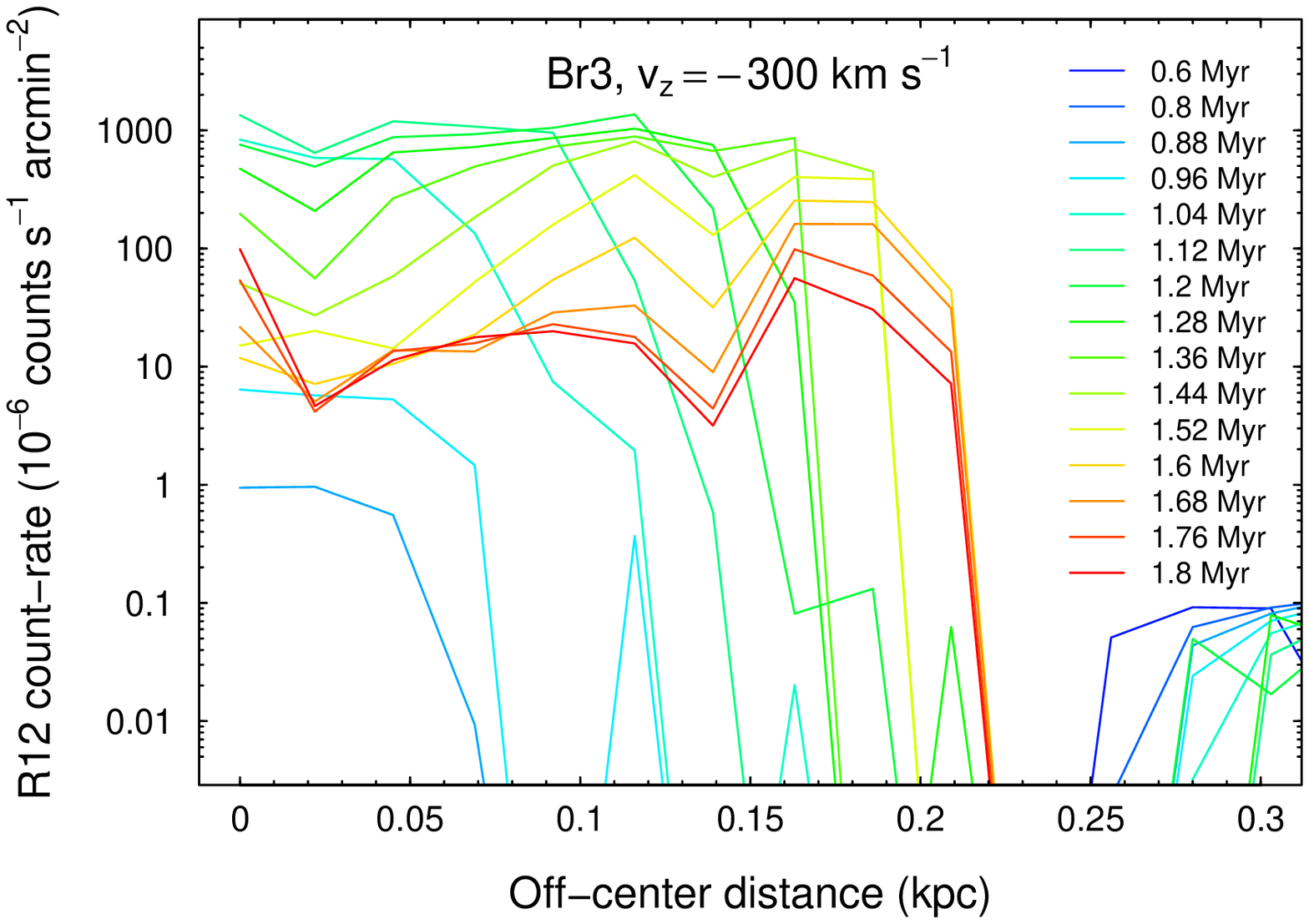}
  \caption{ 
X-ray count rates due to the shock produced by the fast-moving
infalling
clouds in Model \adiabatic 3 (top panel), Model \warmism 3 
(middle panel), and 
the higher resolution Model \nei 3
(bottom panel). 
The \rosat\ R12 count rate, i.e. the count rate in the 
1/4 keV band,
was calculated for vertical sight lines through the
computational domain and is plotted relative
to the distance from the projected center of the cloud. 
The background count rate has been subtracted from all 
of the plots.
Results for Models \warmism\ and \adiabatic\ are shown 
at 2~Myr intervals
until the clouds reach the bottom of their computational
domains.    The time granularity between stored epochs
is much smaller for the higher resolution \nei\ model and
dimmer epochs at the beginning and end of the simulation
are not shown.}
\label{fig:offcentercountrate}
\end{figure}

When radiative cooling is included in the simulation, the
hot gas in front of the cloud sheds most of its
energy via radiation,
only some of which is in the X-ray band.  
Figures~\ref{fig:1/4kevevolution} and
\ref{fig:offcentercountrate} include plots of the resulting 1/4 keV
surface brightnesses from the high resolution Model \nei\ 
simulations.
The average surface brightnesses across footprints 
that are 200~pc in radius 
peak around 3, \ 500, and $2000 \times 10^{-6}$ 
counts s$^{-1}$ arcmin$^{-2}$ in the \rosat\ 1/4 keV band
within 1 to 1$\slantfrac{1}{2}$~Myr of the collision
for the $v_z = 200, 300$ and $400$~km~s$^{-1}$ collisions,
respectively.

The surface brightness profiles of Models \adiabatic 3 and
\warmism 3 in Figure~\ref{fig:offcentercountrate}
show that the surface brightness peaks
near each cloud's axis and drops non-monotonically with radius.
Model \nei 3's profile is more complicated.
The peaks and valleys in the profiles are
associated with the density and temperature structure of
the off-axis gas.
When the gas is very bright, the profile extends beyond the 
footprint of the cloud.  
Sensitive X-ray observations of the clouds would see 
extended disks.

\begin{figure}
\epsscale{1.2}
\plotone{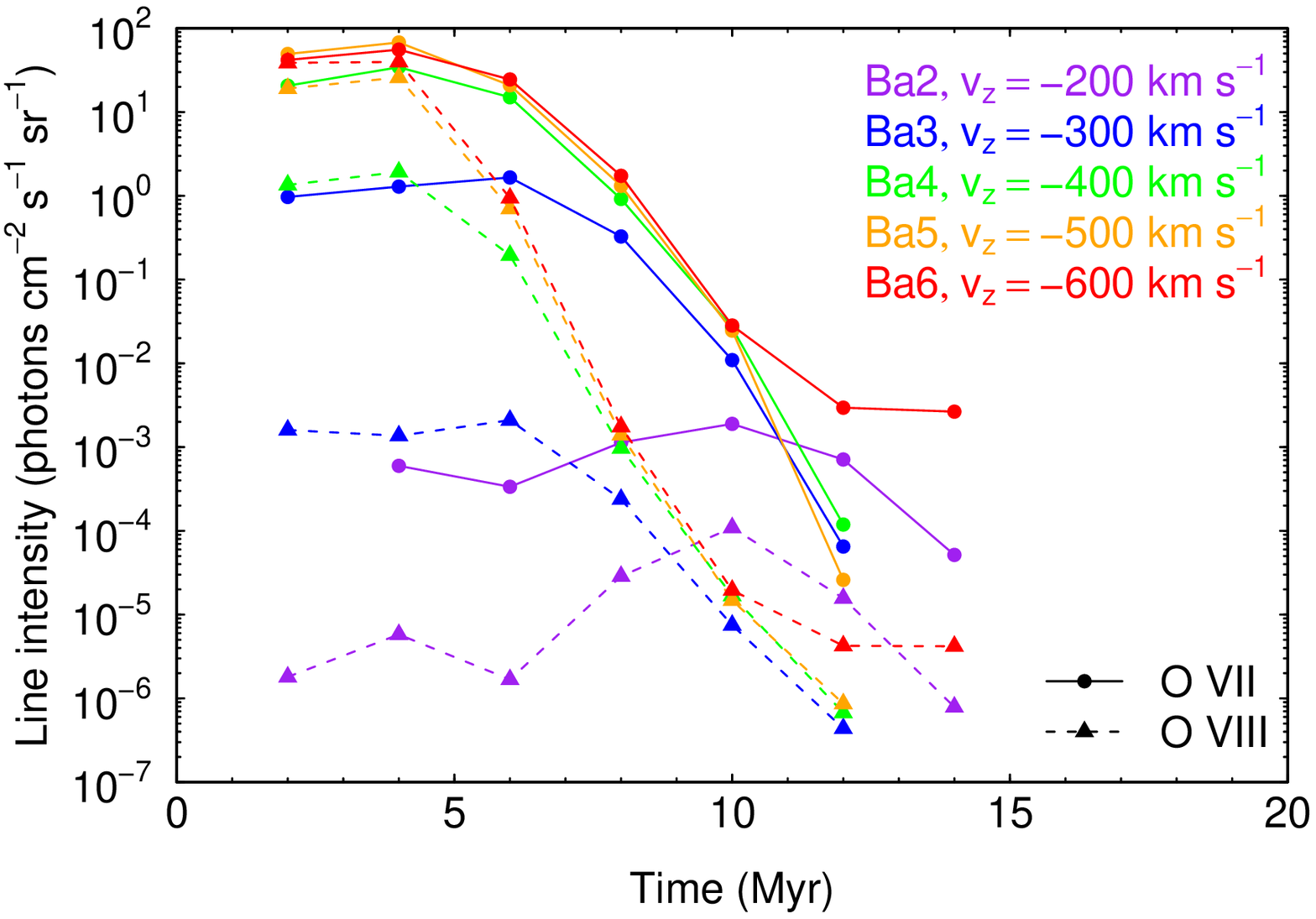}
\plotone{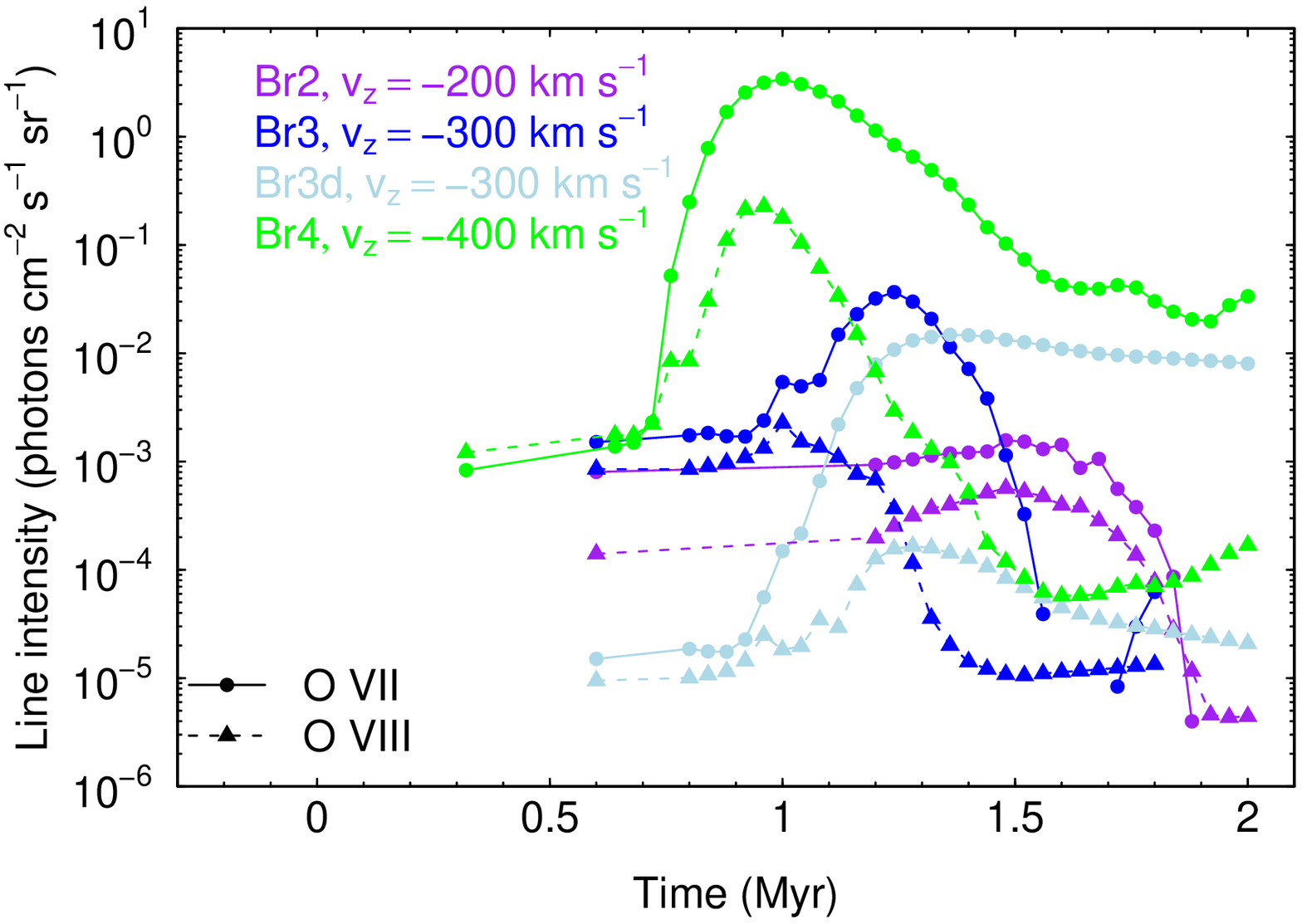}
  \caption{
The \oxyseven\ and \oxyeight\ intensities as functions of time 
for Models \adiabatic 2, \adiabatic 3, \adiabatic 4, 
\adiabatic 5, and \adiabatic 6 (top) and the high
resolution versions of Models \nei 2,
\nei 3, \nei3d, and \nei 4 (bottom). 
The intensities are given in photon units
(photon~s$^{-1}$~cm$^{-2}$~sr$^{-1}$) and averaged across a disk
of radius 200~pc.  The Case \warmism\ models are not shown, 
but are roughly similar in brightness to the Case \adiabatic\ 
models.   As can be seen by comparing
the top and bottom panels, the Model \adiabatic\ peak intensities
are roughly an order of magnitude brighter than the 
Model \nei\ peak intensities. 
\\
}
\label{fig:ovii_oviii}
\end{figure}

The shock-heated gas is less conspicuous in the 3/4 keV
band than in the 1/4 keV band, in both the radiative cooling
and the non-cooling scenarios.
For both Models \adiabatic 3 and \warmism 3, 
the \oxyseven\ triplet 
(photon energy $\sim 570$~eV) intensity
across a disk of radius = 200~pc 
averages to $\sim1$ photon s$^{-1}$ cm$^{-2}$ sr$^{-1}$
for the first several million years (see Figure~\ref{fig:ovii_oviii}).
It averages to far less than 
1 photon s$^{-1}$ cm$^{-2}$ sr$^{-1}$
for Model \nei 3.  The former intensity is
about $\sim1/5$ of the typical observed intensity on 
randomly chosen sight lines away from the Galactic center
\citep{henley_shelton_10}.
Because the collisions weaken with time,
the \oxyseven\ intensities fade
sooner than the 1/4~keV surface brightnesses.
Furthermore, none of the $v_z = -300$~km~s$^{-1}$
models produce appreciable \oxyeight\ intensities at any time.

\begin{figure}
\epsscale{1.2}
\plotone{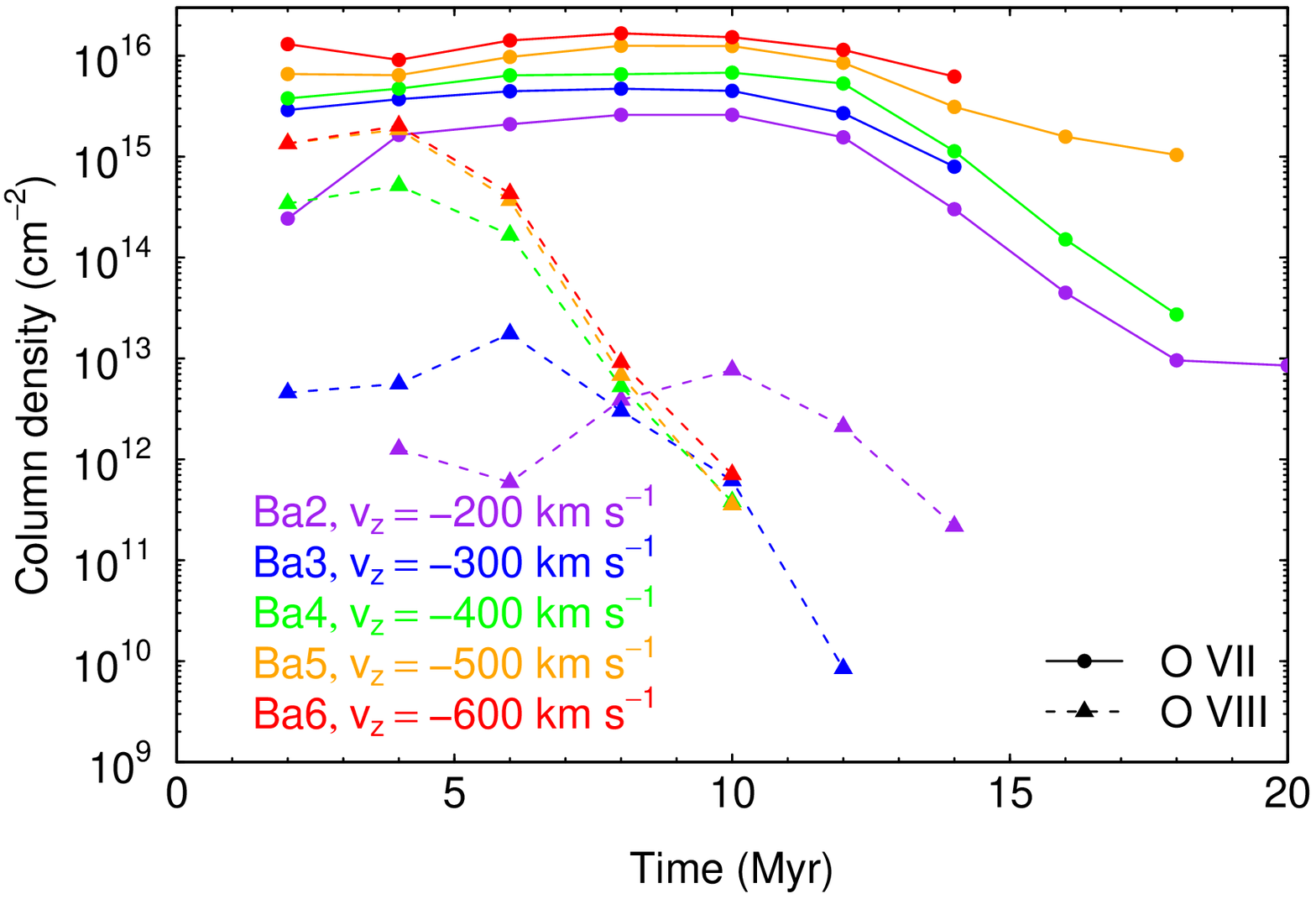}
\plotone{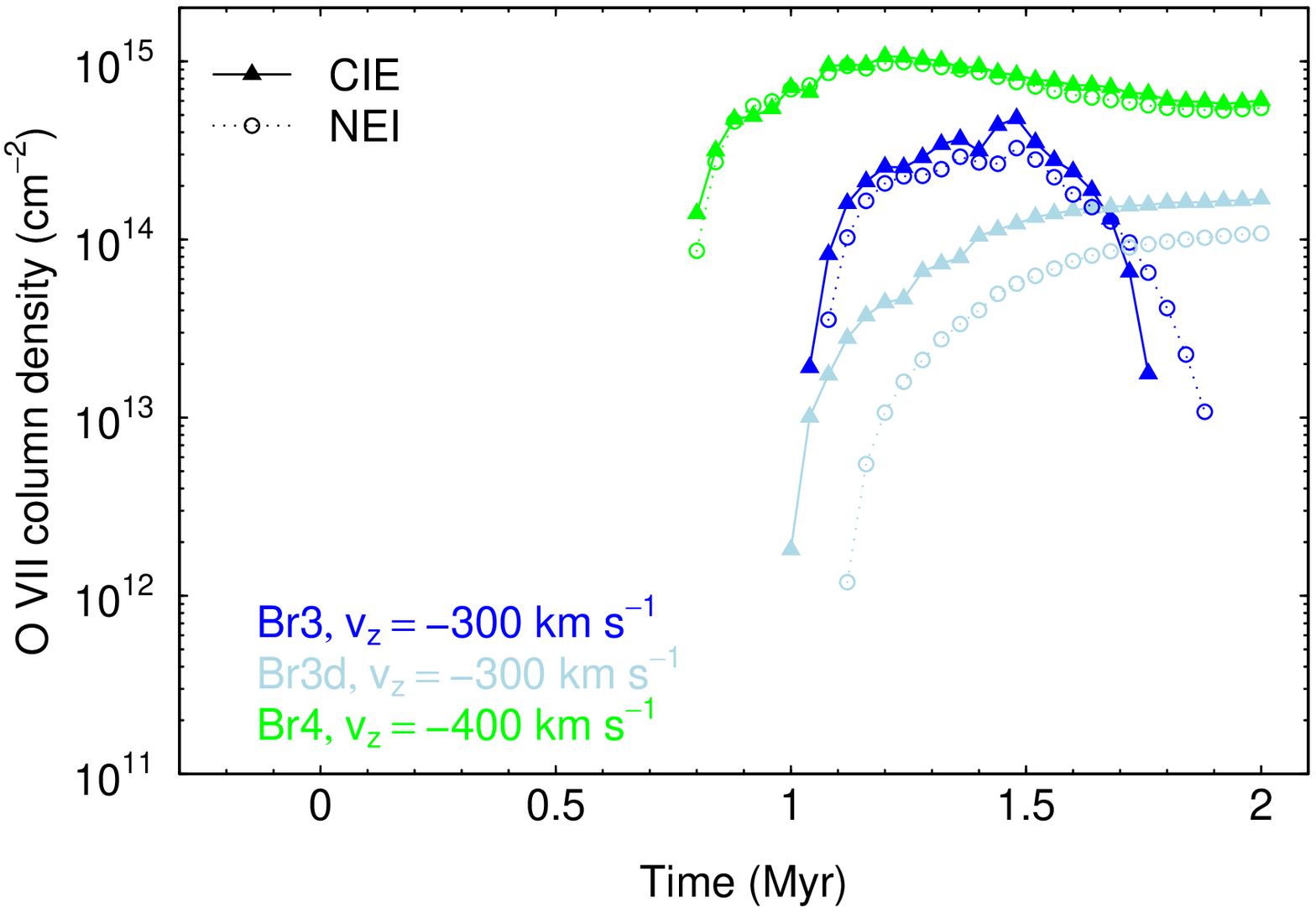}
\plotone{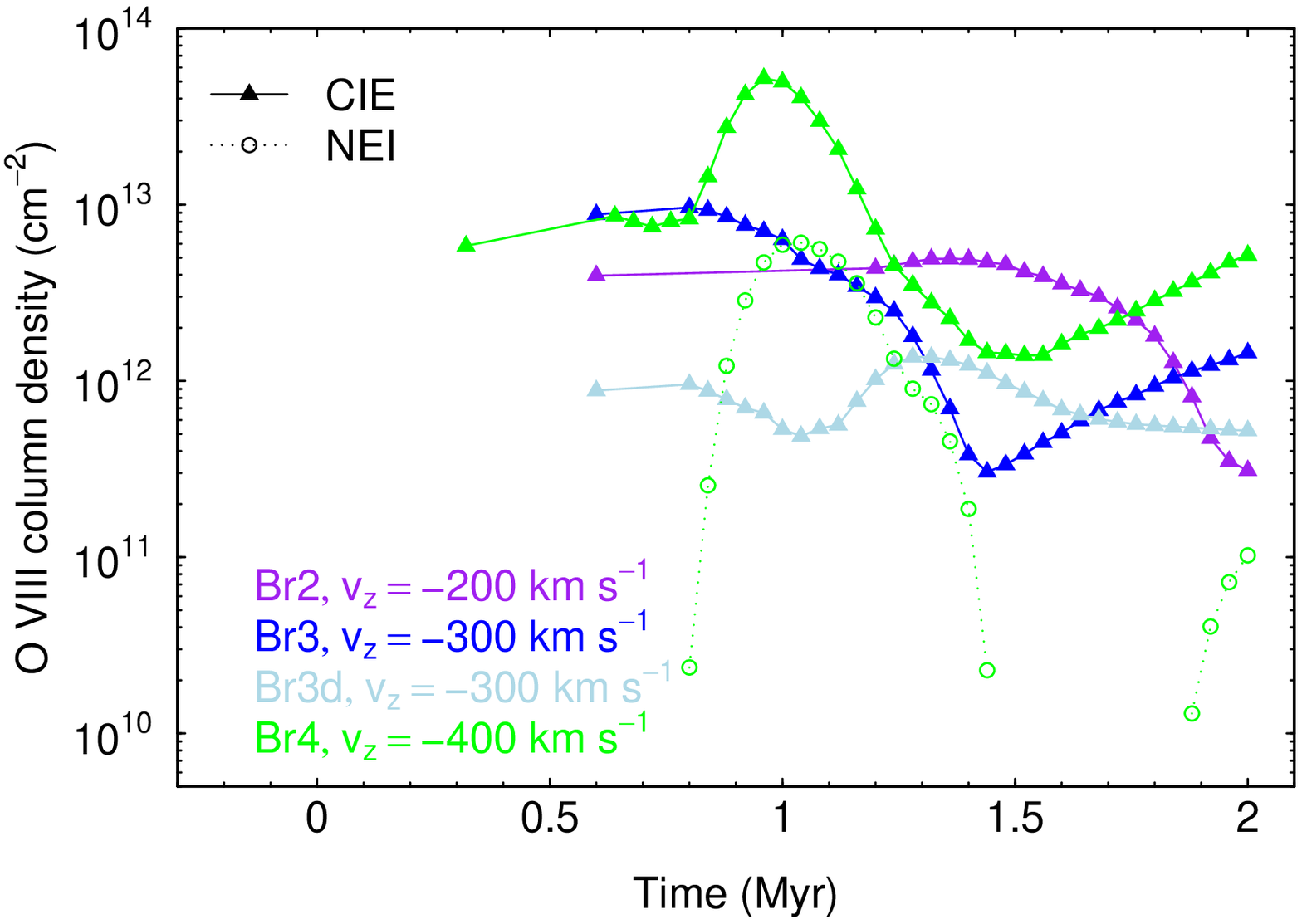}
 \caption{Excess \oxyseven\ and \oxyeight\ column densities as
a function of time for Model \adiabatic\ simulations and 
for high resolution versions of Model \nei\ simulations.    
The hot ambient gas at the top of the simulational domain 
produces a background \oxyseven\ column density of 
$1.94 \times 10^{14}$~cm$^{-2}$ which has been subtracted and so
does not contribute to the plotted values.   
The background-subtracted
column densities have been averaged over disks of radius 200~pc 
in order to obtain the plotted values.
For the \nei\ models, we provide two predictions.   
The dashed lines
mark the column densities calculated using the CIE approximation
while the dotted lines mark the column densities obtained from
NEI calculations done in FLASH.
	This plot demonstrates that the \oxyseven\ 
	column densities are greater for faster cloud
	speeds, greater when radiative cooling is not allowed,
	and remain high later than the \oxyseven\ intensities.
 }
\label{fig:ovii_oviiicolden}
\end{figure}

The average \oxyseven\ column density 
within a footprint of radius 200~pc is 
$3 \times 10^{15}$~cm$^{-2}$,
for both Models \adiabatic 3 and \warmism 3 at 2 Myr.
It remains near this level for 10~Myr,
thus outliving the duration of the bright \oxyseven\
emission.   
In Model \nei 3, the average \oxyseven\ column density
is $\sim4 \times 10^{14}$~cm$^{-2}$ 
from about 1.3 to about 1.5~Myr.
The \oxyseven\ and \oxyeight\ column densities for
these and other speeds simulated for Model \adiabatic\ and
the high resolution versions of Model \nei\ are plotted in 
Figure~\ref{fig:ovii_oviiicolden}.

The gas in these models is also UV emissive.  
When we divide the predicted energy spectra into 0.25-dex wide
bins, we find that the 
10 to 18 eV photon energy range is the most luminous
part of the $\sim5$ to $\sim 2000$~eV spectrum for Models
\adiabatic3, \nei3, and \warmism3.   
Model \nei 3, for example, radiates
more than 100 times more power in 10 to 18 eV photons
than in 100 to 180 eV photons when it is at its peak 1/4 keV
X-ray brightness (age $\sim$1.3~Myr).
At earlier and later times, the ratio is even larger.   
With its comparatively large radiative power, 
the UV plays a more important role in cooling the 
shock-heated gas than does the X-ray.

We have presented results for both radiatively cooling and
adiabatic models.   In the following subsections, we continue
to discuss both cases.
While, on the face of it, it appears contradictory to 
consider the X-ray count rates expected from hot gas whose
hydrodynamics were calculated in models that
excluded radiative cooling (Models \adiabatic\ and
\warmism), we note that 
1.) the true cooling rate is unknown,
2.) if the gas phase abundances are between zero and
the solar values, then the cooling rate should be between
that used in the \nei\ suite of models and those used in
the \adiabatic\ and \warmism\ suites, 
and 3.) X-ray emission accounts for very little cooling.
Because 
X-ray emission accounts for such a small fraction of 
the cooling
and because the 
relative contribution to the spectrum made by any 
given element varies from one energy band to the next,
it is possible for variations in gas phase metallicities to
preferentially lower the cooling rate 
compared with those in Case \adiabatic\ or
raise the X-ray count rate 
compared with those from Case \nei.
Iron provides a good example of the effect of the gas-phase
metallicity.
Almost $60\%$ of the energy radiated in 5 to 100 eV photons
by a $T = 10^6$~K plasma
having \citet{allen_73} abundances are emitted by
iron ions, while slightly less than $20\%$ of the 
\rosat\ R12 counts
are due to photons emitted by these ions.
Thus, if the true iron abundance in the shock-heated gas
were half that expected by
\citet{allen_73}, then the overall radiative cooling rate 
would be noticeably affected (it would be about $70\%$ of 
the CIE rate), while the 
\rosat\ R12 count rate would be insignificantly affected
(it would be about $90\%$ of CIE rate).
Reducing the iron abundance to zero decreases the radiative
cooling rate by 3/5,
while reducing the 1/4 keV luminosity by only 1/5.
Similar phenomena occur at higher temperatures.
Silicon provides another example, because it accounts for
$13\%$ of the 5 to 100 eV power, but almost $30\%$
of the \rosat\ R12 count rate.   Thus, doubling its relative
abundance would increase the radiative loss rate by only
1/8, but increase the 1/4 keV count rate by almost third.
Here, we aren't suggesting that the halo has supersolar
abundances of silicon, but are pointing out that if the
ratio of silicon to other elements is larger than in solar
abundance gas, the effect would preferentially benefit the X-ray
luminosity.   The interested reader is referred to 
\citet{sutherland_dopita_93} figure 18 for relative cooling
rates from various elements.
Other reasons why the rate of radiative cooling might 
differ from the CIE rate for solar abundances include
disequilibrium between electron and ion temperatures and
delayed ionization and recombination.
\\

\subsubsection{Effect of HVC Velocity}
\label{subsect:velocity}

We ran several additional simulations in order to explore the 
effects of initial velocities.  These, plus the simulations 
already mentioned, create several suites.
Models \adiabatic 2, 3, 4, 5, and 6 
have initial cloud velocities of $v_z = -200, -300, -400, -500$,
and $-600$~km~s$^{-1}$, respectively.
Models \warmism 3, 4, 5, and 6 have velocities of 
$v_z = -300, -400, -500$, and $-600$~km~s$^{-1}$, respectively.
The faster end of this range is greater 
than expected for
HVCs near our galaxy, but may be of value for studying 
more energetic collisions.   Models \nei2, 3, and 4
have velocities of
$v_z = -200, -300$, and $-400$~km~s$^{-1}$, respectively.
We simulated the Model \nei\ suite in both moderate and high 
resolution forms, but for this analysis, we focus on only the 
high resolution simulations.

Even the $v_z = -200$~km~s$^{-1}$ cloud induces bright X-rays
in the non radiatively cooled simulations and some
X-rays in the radiatively cooled simulations.
These X-rays are from gas that was shock-heated to 
$T \sim5 \times 10^5$~K.
The faster clouds shock-heat the ISM to higher temperatures.
As a result, they produce more 1/4 keV X-rays for longer
periods of time 
(see Figure~\ref{fig:1/4kevevolution}).
Furthermore, the \oxyseven\ and \oxyeight\ intensities, 
which are modest in models \adiabatic 2 and \adiabatic 3,
become stronger in Models \adiabatic 4, 5, and 6 
(see Figure~\ref{fig:ovii_oviii}).
Because the shock heated gas is more highly ionized 
in the faster models, the column densities of 
\oxyseven\ and \oxyeight\ are also larger 
(see Figure~\ref{fig:ovii_oviiicolden}).
\\

\subsubsection{Effect of Density}
\label{subsect:density}

The density of the environmental gas affects the hydrodynamics 
and production of 1/4~keV X-rays in multiple ways.
When the ambient density is greater, then the swept up, 
shock-heated environmental gas in front of the cloud has greater
density and depth, resulting in a higher 1/4 keV surface 
brightness.
In addition, denser gas better decelerates the cloud, resulting 
in lower post-shock temperatures.   The effect of lowering
the post-shock temperature, however, can increase or decrease
the emissivity, depending upon the circumstances.
Given that the peak in the theoretical 1/4 keV X-ray 
emissivity curve
is near $T = 0.9 \times 10^6$~K for CIE plasmas, slowing the cloud
increases the 1/4~keV X-ray emissivity in the cases of 
very fast clouds (e.g., those with initial 
$v_z = -600$~km~s$^{-1}$) and
decreases it in the cases of only moderately fast clouds
(e.g., those that have reached $v_z = -250$~km~s$^{-1}$.)
With each of these factors playing a role, our Model \warmism 3 
is 10 times brighter than its less dense counterpart, 
Model \warmism 3d, whose environmental density is 1/10 that
of Model \warmism 3.  Our Model \warmism 6 is 100 times brighter 
than its less dense counterpart, Model \warmism 6d,
whose environmental density is 1/10 that
of Model \warmism 6,  
and our Model \nei 3 is $\sim30$ times brighter
than its less dense counterpart, Model \nei 3d,
whose environmental density is 1/10 that
of Model \nei 3.
Furthermore, the greater the amount of shock-heated
interstellar material that is swept
up, the greater the column densities of very high ions.
Consequently, Models \warmism 3 and \warmism 6 have about 10 times
greater \oxyseven\ column densities
than Models \warmism 3d and \warmism 6d, respectively, while
Model \nei 3 has roughly twice the \oxyseven\ column 
densities of Model \nei 3d.
\\

\subsection{Turbulent Mixing}
\label{subsect:hotism}

Here, we consider mixing between cool and warm clouds 
as they travel through hot, rarefied ambient gas.
As the cloud passes through the hot gas, 
Kelvin-Helmholtz instabilities develop and
mixing between the cloud and ambient material creates 
a zone of intermediate temperature, intermediate
density gas \citep{esquivel_etal_06}.   
Analytical calculations
and computational simulations
have already shown that mixed layers are rich in
high ions, such as \oxysix\ 
\citep{slavin_etal_93,esquivel_etal_06,kwak_shelton_10,kwak_etal_11}.
In order to determine whether or not mixing zones between
HVCs and very hot gas can also be X-ray productive,
we simulate the hydrodynamics of 3 dimensional clouds 
passing through hot, rarefied gas.
This approach differs from the more common approach, which
treats the two regions as blocks that slide past each other,
but better replicates the effects of the cloud's 
rounded shape and motion.    The names of these simulations
begin with \hot.

\begin{figure*}
\epsscale{1.0}
\centering
\includegraphics[scale=0.475]{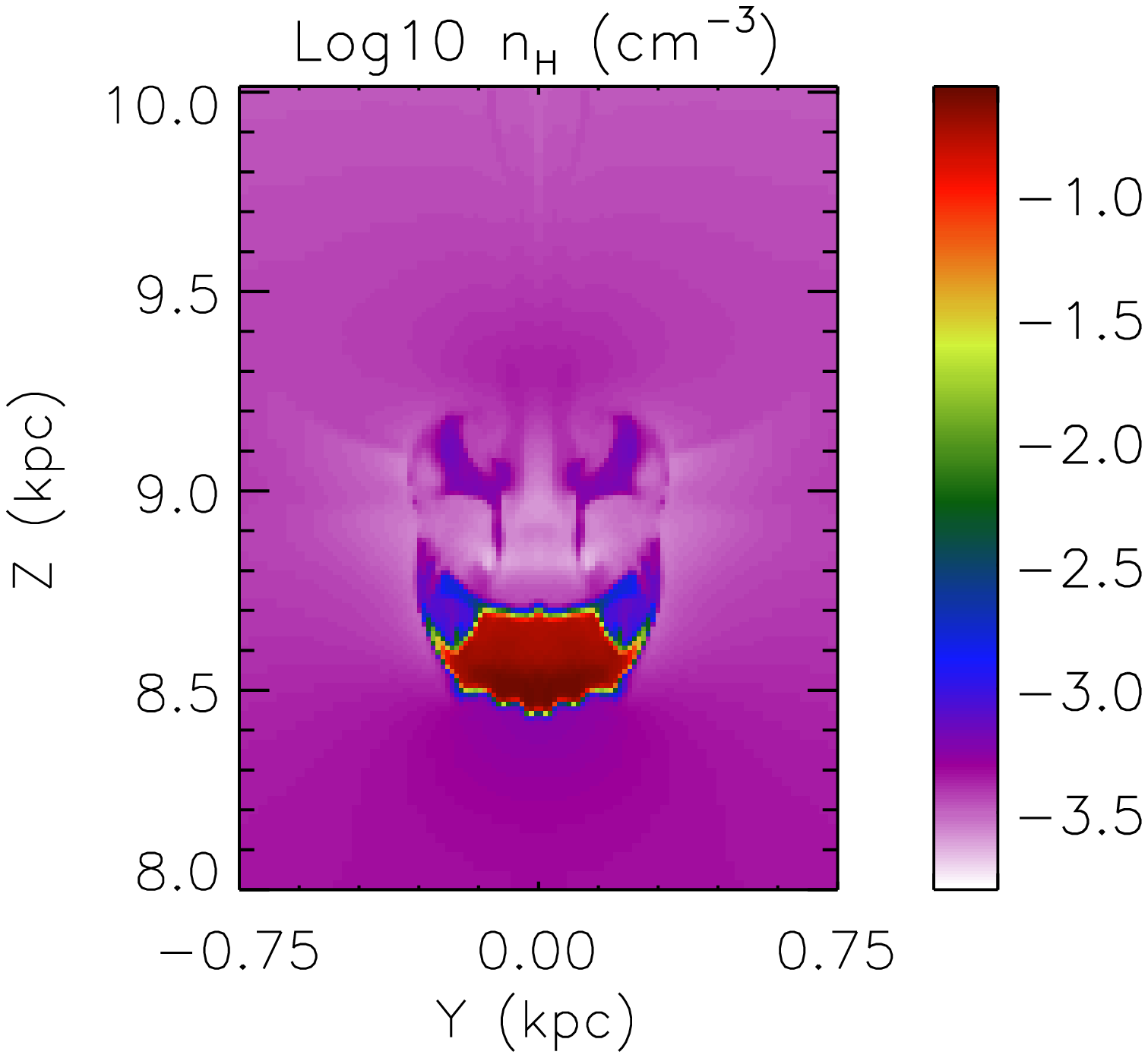}
\hspace{1.25in}
\includegraphics[scale=0.475]{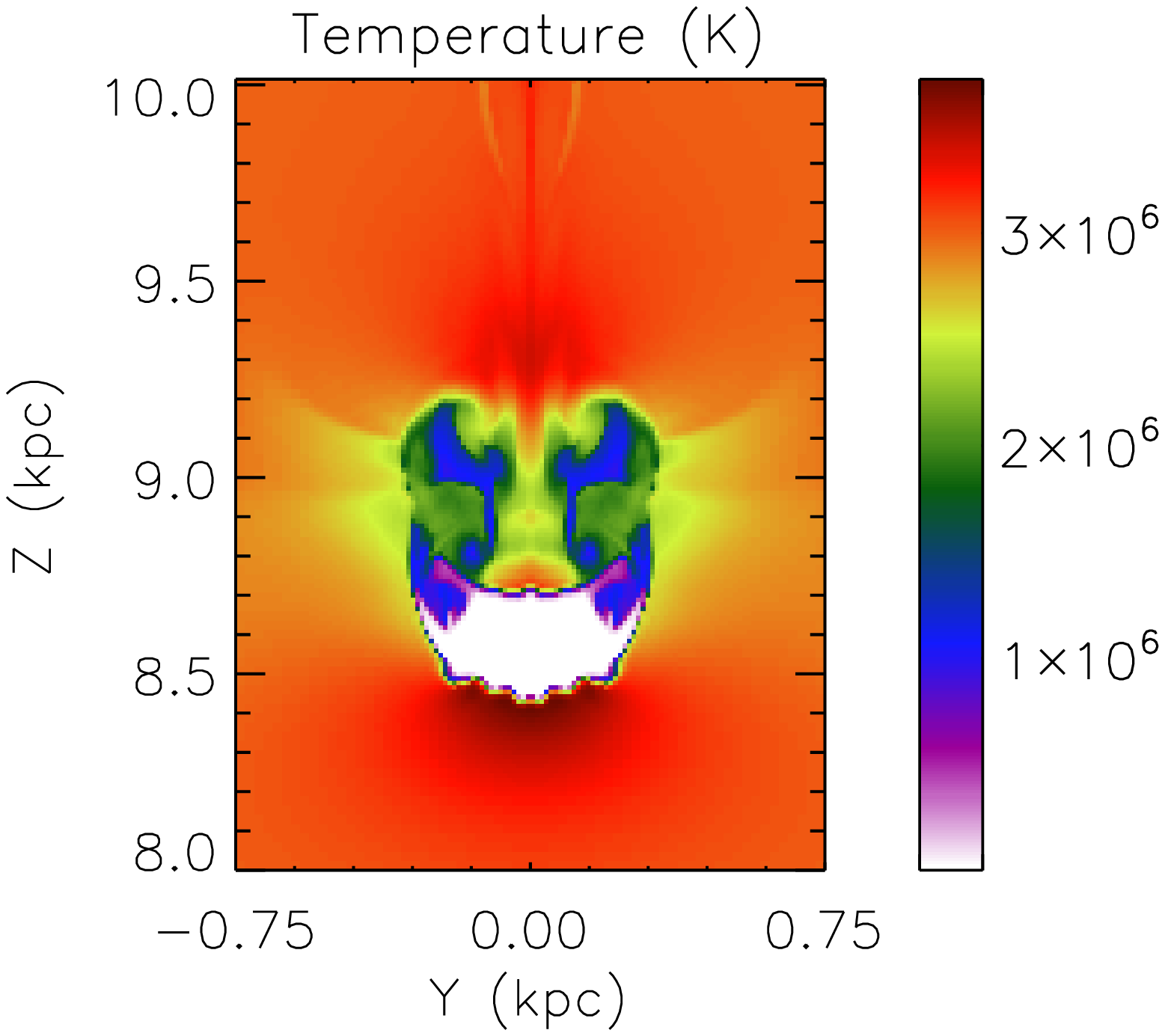} \\
\vspace{0.75in}
\includegraphics[scale=0.475]{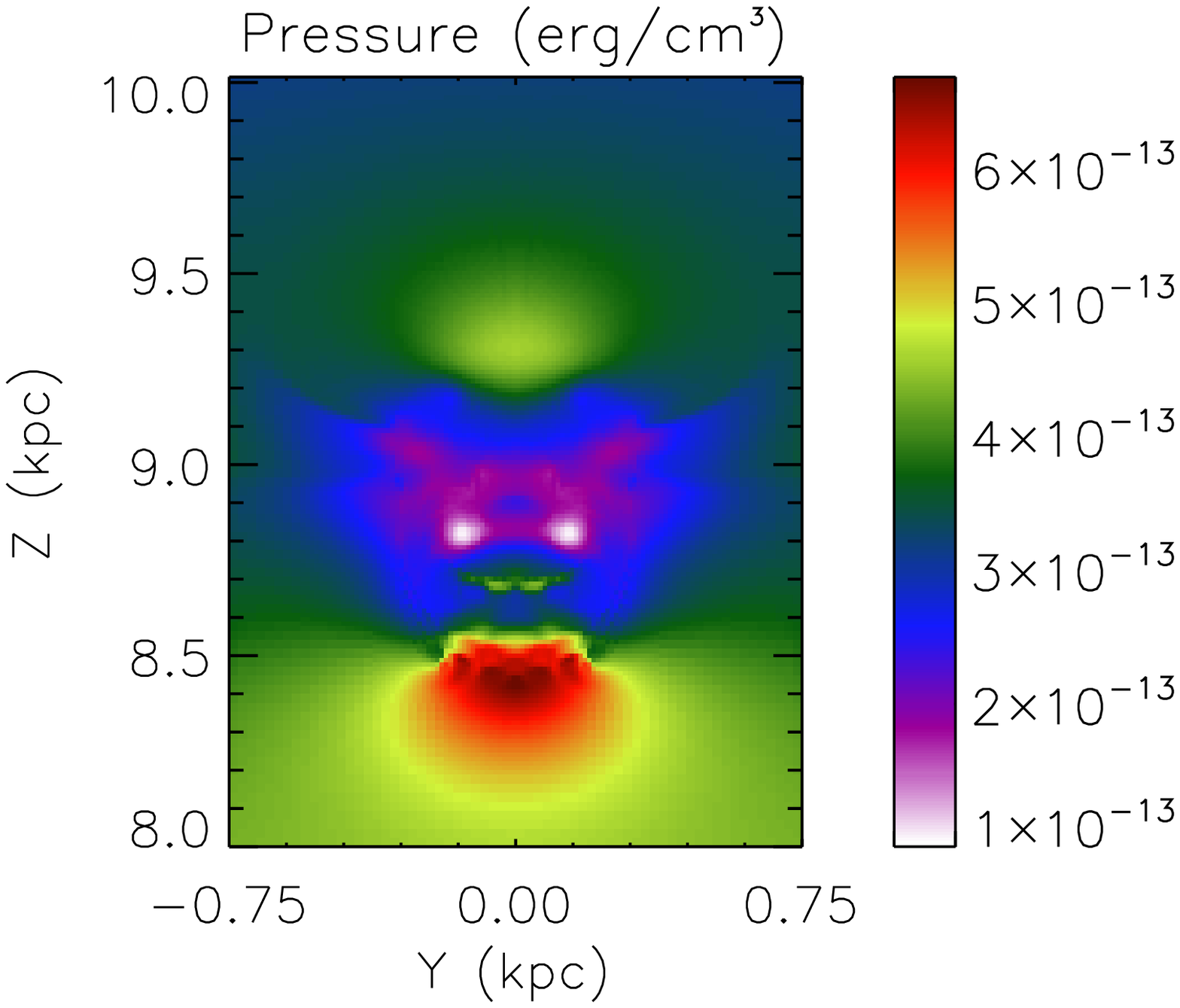}
\hspace{1.25in}
\includegraphics[scale=0.475]{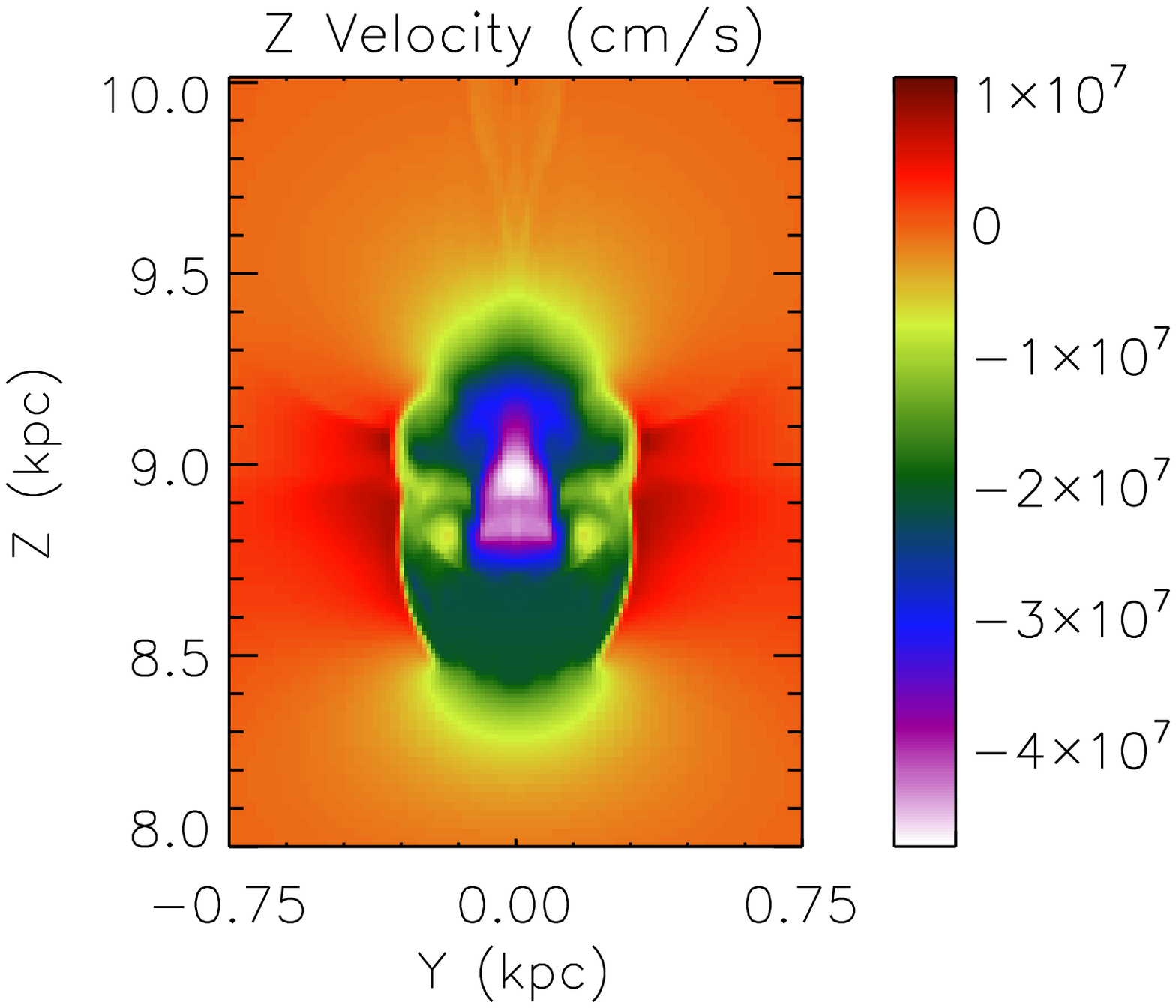}
\vspace{0.5in}
  \caption{  
Number density of hydrogen atoms (top left), 
temperature (top right),
pressure (bottom left), and velocity in the $z$ direction (bottom right)
of the \hot11 cloud and the material ablated from it
after the cloud has fallen for 30~Myr through hot halo gas.   
The density plot is logarithmic, while the others are linear.
The images show a vertical slice through the 
center of the cloud at $x = 0$~kpc.
The ablated and mixed material appear to form ears on the sides of the
cloud in the top two figures.   
Because of the symmetry in the initial conditions,
the trailing stream of dislodged gas actually forms
a crown shape.   The pressure behind (i.e., above) the 
cloud is noticeably low. 
The apparent pixelation is an artifact of the image processing and
does not represent the resolution of the simulations.
%
\\
}
\label{fig:cloudwithears}
\end{figure*}

In order to model the interaction adequately, the
simulations must resolve the mixing length scale.
Interstellar turbulence may exist across 
a wide range of length scales, but 
the largest length scale dominates the mixing
\citep{esquivel_etal_06}.
In our case, 
the largest length scale is the height of the distorted
cloud.  This height and width are adequately resolved in our 
Model \hot\ simulations,
which, when fully resolved, use $\sim30$ zones to model a 
400~pc span (the nominal height of the
interaction zone in Figure~\ref{fig:cloudwithears}).

Our simulations do not include thermal conduction, but
\citet{deavillez_breitschwerdt_07}
found that turbulent diffusion is much
more effective than thermal conduction 
at leveling the temperature gradient.  Thus,
our omission of thermal conduction is not problematic.
Note, also, that radiative cooling has not been allowed 
in the Model \hot\ simulations.   It was disallowed
in order to maintain hydrostatic balance in
the background gas.
The ramifications of cooling are too
complicated for us to simply say that adding cooling
would lower the radiation rates of the turbulently mixed
gas.   This is
because radiative cooling would not only allow 
the $\sim1 \times 10^6$~K gas to cool to
lower temperatures, but it would allow hotter gas 
to cool to $\sim1 \times 10^6$~K.

Model \hot1\ serves as an example of Case \hot\ simulations
and as our reference simulation for comparison with the others.
In this simulation, we
track the cloud as it falls from its initial
height of 12~kpc to near the bottom of our grid
($z = 8$~kpc), which takes 32~Myr of simulated time.
We record the simulation results at 2~Myr intervals and
here describe the results at two of the sampled epochs.
By 16~Myr, the cloud has accelerated to $v_z = -120$~km~s$^{-1}$ 
and fallen to a height of $z \sim 11$~kpc.
By 30~Myr, it has accelerated to $v_z = -220$~km~s$^{-1}$ 
and fallen to $z \sim 8.2$~kpc.
We obtain the cloud-induced 
excess intrinsic count rate in the 1/4~keV band
by subtracting the count rate of the
undisturbed hot halo from that toward the disturbed gas.
Vertical sight lines throughout the domain are sampled.
At 16~Myr, the brightest X-ray excess is 
$0.1 \times 10^{-6}$~counts~s$^{-1}$~arcmin$^{-2}$.   
At 30~Myr, it is $0.4 \times 10^{-6}$~counts~s$^{-1}$~arcmin$^{-2}$.
The latter is the brightest rate for Model \hot 1
at any time.
Note that these are intrinsic count rates;
absorption by $N_H \sim 10^{20}$ cm$^{-2}$ would
decrease them by factors of $> 2$.    
Even without absorption,
the small count rate enhancements would be inconspicuous
against the diffuse X-ray background.   

The timescale for the growth of the Kelvin-Helmholtz instabilities
decreases as the cloud falls and the density contrast between the
cloud and the ambient gas decreases. For the last $\sim14$~Myr of the
simulation, we estimate that the timescale for the growth of
instabilities is $\la14$~Myr, from
\citet{chandrasekhar_61} Section 101, and
assuming that instabilities grow on similar timescales in 
plane parallel geometries as in our geometry.  
Hence, the lack of X-rays due to turbulent mixing in our simulations is
unlikely to be due to the instabilities having insufficient time to
grow.

Models that simulate
a larger ambient temperature 
(Models \hot4 and \hot11),  
a greater ambient pressure (Models \hot5 and \hot11), 
faster initial speeds 
(Models \hot2, \hot8, \hot9, and \hot10),
denser clouds (Model \hot11),
less dense clouds (Models \hot3 and \hot10),
and nonzero magnetic fields (Models \hot6 and \hot7)
were also simulated.
Of the slow models (initial $|v_z| \le 50$~km~s$^{-1}$) 
the greatest X-ray enhancement
seen anywhere in the domain at any time during the
simulations is 
$35 \times 10^{-6}$ counts~s$^{-1}$~arcmin$^{-2}$.
This occurred in Model~\hot11 at 30~Myr, just before 
the cloud crossed the domain's lower boundary.
The next brightest model was Model \hot4, with 
a peak enhancement of
$14 \times 10^{-6}$ counts~s$^{-1}$~arcmin$^{-2}$
at 30~Myr, also just before 
the cloud crossed the domain's lower boundary.
Models \hot4 and 11 have the largest ambient gas temperature 
($T_{ISM} = 3 \times 10^6$~K) 
and, through most of the domain, have the largest
ambient density of all of the Case \hot\ models.
These attributes prime Models \hot 4 and 11 to be bright;
the hotter ambient medium produces hotter mixed
gas, while the denser ambient and cloud gas produce
denser mixed gas than the other models.
Although Model \hot11 is within a factor of a few of being 
comparable with X-ray observations, suggesting that 
further adjustment of the model parameters could 
result in a sufficiently bright model,
the parameters in
Model \hot11 are already near reasonable limits.

Some of the fast models 
(initial $|v_z| = 300$ or 400~km~s$^{-1}$)
create brighter disturbances.
Model \hot9, for example, has a peak enhancement of
$78 \times 10^{-6}$ counts~s$^{-1}$~arcmin$^{-2}$
at 8~Myr, which is the last epoch before the cloud leaves the domain.
The bright region is fairly extended, with a radius
of 740~pc (with the cut off being defined where the
count rate exceeds the background by 5$\%$)
over which the count rate averages
$75 \times 10^{-6}$ counts~s$^{-1}$~arcmin$^{-2}$.
While an excess of this magnitude and spatial extent may
be observable, it is not due to turbulent mixing.
It is due to the shock. 
Likewise, the shock in Model~\hot8 
created a bright region, whose maximum average brightness was 
$20 \times 10^{-6}$ counts~s$^{-1}$~arcmin$^{-2}$
over a 200~pc radius footprint at 10~Myr, the last epoch before
the cloud left the grid.
Again, the X-rays are due to the shock, not the mixed material.

Here we consider the dynamics of the mixed gas and how they lead to
X-ray dimness.    
Although our models develop a mixed zone, as expected, 
and although the mixed zone contains some hot gas,
in several models (\hot1 - \hot3 and \hot5 - \hot10)
too little of the mixed gas was
sufficiently hot (i.e., $T$ nearing $10^6$~K) to be X-ray emissive
and even when the temperature was sufficiently high,
as it was in Models \hot4 and \hot11, 
the mixed gas fell behind the clouds into the semi-vacuum
created by the cloud's passage.   Here the density was not
sufficient for great emissivity.
Model \hot11 provides an example of the varying conditions and
low pressure in the ablated material 
(see Figure~\ref{fig:cloudwithears}).
Within the trailing stream the temperature and density
vary from hot ($T = 2.6 \times 10^6$~K) but
diffuse (stream density $\sim90\%$ of the density in the 
undisturbed halo gas at this height) near the surface of the stream
to only mildly hot ($T = 1.2 \times 10^6$~K) but denser 
(stream density $\sim 180\%$ of the density in the undisturbed gas)
in the core.  
The pressures in these example locations
are $\sim70\%$ those of the ambient gas.

The enhancement due to turbulent mixing by a single cloud is modest
when compared with the typical X-ray count rate
for high latitude sight lines.
Scattered clouds may contribute unidentifiably to the
soft X-ray background and multiple, aligned clouds
could create a non-negligible X-ray surface brightness.
However, obtaining bright X-rays from individual clouds
requires fast speeds as discussed in the preceding section.

     Like the suites of \adiabatic, \nei, and \warmism\ models, 
the suite of A models
radiates more prolifically in the UV than in the X-ray.   The
broadband spectra are several orders of magnitude brighter in the
far UV than in the soft X-ray.    The ultimate source of this energy
is the reservoir of thermal energy in the hot ambient gas.  Mixing 
with the ablated cloud material has lowered the temperature and 
ionization level of the neighboring hot gas such that it has become 
highly emissive, especially in the UV.
\\

\subsection{Observational Appearance}
\label{subsect:obsappearance}

To the observer
who looks straight upwards at an incoming HVC, the X-ray 
bright region would extend for at least 200~pc from the
center of the cloud, thus 400~pc in diameter.   
If the observer is not located directly
beneath the cloud, then the bright region would be somewhat
displaced from the cloud itself and would subtend a smaller
angle in the direction along the cloud's motion.
If the clouds were as far away as Complex C, 
which is located $\sim12$~kpc from Earth, then
the 400~pc diameter X-ray bright footprint
due to a single cloudlet would
subtend only a 2\degr\ angle.   Such a small feature 
cannot be easily examined with \rosat\ All Sky Survey (RASS) data.
However, the overlapping footprints of many bright clouds in an 
ensemble, could create an extended
X-ray bright region that should be compared with the RASS maps.

If Complex C is composed of such an ensemble, then we can estimate
the average 1/4~keV count rate across the complex.
Here we examine the cases in which the X-rays result from 
shock heating by assuming that the individual clouds within 
Complex C
are like Model \adiabatic 3 or \nei 3 clouds; faster models would 
result in greater X-ray count rates while slower models 
or more 
cooling would result in lower count rates.  Any gas that was
ablated from the clouds and mixed with the ambient medium
would also be dim. 

The expected count rate is a product of
the count rate of a single cloud when viewed
from below ($r$), 
the number of such clouds ($\mathcal{N}$), 
and a scaling factor ($f$) that accounts for the dilution of the
surface brightness over the larger area of Complex C.
For Model \adiabatic 3, at its  brightest epoch,
the average 1/4~keV surface brightness, $r$, within
a circular extraction region
of radius 400~pc (this is twice as large as the previously
mentioned extraction region, in order to capture more of
the photons) is 
3600 $\times 10^{-6}$ counts s$^{-1}$ arcmin$^{-2}$ in the
\rosat\ R12 band.
In model \nei 3, the emission is dimmer, but more concentrated.
The surface brightness in the \rosat\ R12 band for a 200 
pc radius footprint is 
560 $\times 10^{-6}$ counts s$^{-1}$ arcmin$^{-2}$.
The number of individual clouds is the ratio of Complex C's mass 
($M_{CC} = 8.2_{-2.6}^{+4.6} \times 10^6 $~M$_{\odot}$, 
\citealt{thom_etal_08})
to the initial mass of one Model \adiabatic 3 or \nei 3 cloud 
($M = 7.5 \times 10^5$~M$_{\odot}$),
thus $\mathcal{N}$ is $10.9^{+6.1}_{-3.5}$.
If the area of each model extraction region,
$A = \pi \times (400$~pc)$^{2}$ for Model \adiabatic3
and $A = \pi \times (200$~pc)$^{2}$ for Model \nei3, 
were to be diluted so as to
encompass Complex C, whose area is 
$A_{CC} = 3$~kpc $\times$~15 kpc \citep{thom_etal_08}, 
then conservation of luminosity would require 
the average count rate of each 
cloud to be reduced by a factor of $f = A/A_{CC}$.
Not only does this factor account for dilution, 
but it also accounts for the concentration of the X-rays
when the viewing angle results in a foreshortened cross section.
Combining $r$, $\mathcal{N}$, and $f$ yields a theoretical
intrinsic R12 count rate of 
$440^{+240}_{-140}$ counts s$^{-1}$ arcmin$^{-2}$ for the
case in which the clouds are like those in Model \adiabatic 3 and
$17.0^{+9.6}_{-5.4} \times 10^{-6}$ 
counts s$^{-1}$ arcmin$^{-2}$ 
when they are like Model \nei 3.

Attenuation by intervening material will reduce the count rate.
Assuming that $N_H \sim 10^{20}$~cm$^{-2}$,
	which is the typical column density of galactic gas 
	in the directions toward the brighter parts of Complex C,
	$\sim2/3$ to $\sim3/4$ of the original photons 
	will be absorbed or
	scattered before reaching the observer.
	Thus, the observer would see
	an average cloud-induced X-ray surface brightness of
	$\sim 130 \times 10^{-6}$ counts s$^{-1}$ arcmin$^{-2}$
	from a complex of Model \adiabatic 3 clouds.
	This is a significant enhancement, though not
	as large as that seen along some directions in the
	region of Complex C (see Figure~\ref{fig:complexcmap}).
	For a complex of Model \nei 3 clouds, the predicted
	count rate is 
	$4 \times 10^{-6}$ counts s$^{-1}$ arcmin$^{-2}$,
	which is negligible.
	Irrespective of the cooling rate, faster clouds 
	would result in brighter X-rays.


The Magellanic Stream provides another point of comparison.
\citet{bregman_etal_09} report an enhancement of
0.4 to 1.0 keV X-rays on the leading side of the
MS30.7-81.4-118 cloud within
the Magellanic Stream.
With the \xmmnewton\ pn detector, they found an excess of 
$0.64 \pm 0.10$ counts~ks$^{-1}$~arcmin$^{-2}$.
This is a $6.4\sigma$ effect.  
(\citealt{bregman_etal_09}
also found an excess in their \chandra\ data, but at the $1.6\sigma$
level.)
The bright region is
roughly 6 arcmin across, equating to roughly 90~pc in width.
Here, we compare with the count rates for 100~pc wide circular
footprints from  
our Models \adiabatic 3, \adiabatic 4, \nei 3 and \nei 4,
which, with velocities of 300 and 400 km~s$^{-1}$,
bracket the stream velocity (roughly 380 km~s$^{-1}$,
\citealt{blandhawthorn_etal_07}).
Our non-radiative models, Models \adiabatic 3 and \adiabatic 4, 
produce 0.25 and 3.2 
counts~ks$^{-1}$~arcmin$^{-2}$ at their brightest
epochs (6 and 4~Myr, respectively), thus bracketing the observed
value.   Meanwhile, our radiative models,
Models \nei 3 and \nei 4
produce only $7.0 \times 10^{-3}$ 
and 
0.42 counts~ks$^{-1}$~arcmin$^{-2}$ 
at their brightest epochs (1.16 and 0.92~Myr, respectively).
Only the faster non-radiative model is within range of the
Magellanic Stream observations.
Thus, collisions between fast HVC gas and
relatively dense warm gas can account for the observed
X-rays, but do so more easily if the radiation rate is 
quenched and/or the cloud is moving very fast.

As shown in Figure~\ref{fig:ovii_oviiicolden},
our faster Model \adiabatic\ simulations yield \oxyseven\ column
densities of $\gtrsim 4 \times 10^{15}$~cm$^{-2}$.
This is similar to the median column density for sight lines through
the Galactic halo (\citealt{lei_10}, excluding sight lines through
the Galactic Center soft X-ray enhancement), but is also of the
same order of magnitude as the sight line-to-sight line variation
in observed values.   Thus,
in the absence of radiative cooling, HVC-shocked
gas might be observable with future, high resolution
instrumentation that could distinguish fast-moving from slow-moving
ions.   Again, if the shock-heated gas cooled at the CIE rate,
then far fewer \oxyseven\ ions would result, making the region 
unobservable.
\\

\subsection{Resolution Experiments}
\label{subsect:resolution}

In order to examine the effects of computational resolution we
calculate additional versions of Models \adiabatic 3, 4, 5, and 6
and Model \nei 3
using lesser and greater numbers of refinement levels than in
the foregoing simulations, 
i.e., in FLASH, we use lrefine$\_$max = 4 and 6, 
for the additional Model \adiabatic 3, 4, 5, and 6 simulations,
rather than the value of 5 used in the primary simulations
discussed in previous sections of the paper.
We follow a similar pattern for comparison with the 
moderate resolution Model \nei 3 simulations, but also
use lrefine$\_$max = 5 and 6 when making simulations for
comparison with the high resolution  Model \nei 3 simulations,
which used lrefine$\_$max = 7.
We find that within this range, the refinement
level does not affect the timescale on which the clouds fragment,
although it does affect 
the shapes of the clouds and of the X-ray bright regions.
In order to compare the X-ray productivities of the various
simulations, we extract the 1/4 keV X-ray count rates 
within circular regions of radius equal to 400~pc 
for the Case \adiabatic\ simulations and 200~pc for
the comparisons with the Case \nei\ simulations.
(We set the footprints for the Case \adiabatic\ simulations 
to be greater 
than those used in earlier parts of this paper
in order to capture all or nearly all of the downward directed 
flux.)  We extract these 
count rates from every epoch in the Model \adiabatic 4-like 
simulations,
every epoch from the Model \nei 3-like simulations that were
made for comparison with the primary moderate resolution 
Model \nei 3 simulation, 
every epoch from the simulations that were made for comparison
with the high resolution Model \nei 2, 3, and 4 simulations, 
and the 
$t = 10$~Myr epoch from the Model \adiabatic 3, 4, 5, and 6-like 
simulations.  We then compare our resolution experiment 
simulations with the control simulations
having the same initial cloud velocity.
The X-ray count rates vary somewhat between our test
simulations, but in almost all cases are
within $45\%$ of that of the relevant
control simulation. 
Frequently, they are much closer to those
of the control simulations. 

We also examined 
2-dimensional cylindrically symmetric simulations using
maximum refinement levels of 4, 5, 6, 7, and 8.   The
principle advantage of this configuration is that it allows
much smaller zone sizes.   The principle disadvantage is
that 2-d simulations
are not able to track azimuthal instabilities
\citep{korycansky_etal_02}.
Lacking azimuthal modulation, the cloud material 
concentrates along the symmetry axis and resists fragmentation
until later times 
than in the 3-D simulations.   
In general, the 2-dimensional simulations predict similar X-ray count rates
as the 3-dimensional simulations, but with greater
variation between runs having different refinement levels.
\\

\section{Summary and Discussion}
\label{sect:discussion}

We performed a series of hydrodynamic simulations aimed at 
understanding the X-ray productivity of HVCs.   We examined two 
types of interactions, shock heating and mixing between cloud and 
ambient gas.  We also examined the effect of using CIE ionization 
levels on the X-ray count rates, finding
the 1/4 keV count rates to be similar to those in which we tracked 
the ionization levels in a time dependent fashion, and we examined 
the effect of radiative cooling in the shock scenario, 
finding it to be very important.

We found that shock heating is far more effective at inducing X-ray
emission than turbulent mixing.    Clouds with sufficiently fast
initial speeds ($\ga 300$~km~s$^{-1}$) shock heat the ambient
gas to temperatures of $\ga 1 \times 10^6$~K and clouds with 
initial
speeds of $\ga 400$~km~s$^{-1}$ shock heat the ambient gas
to $\ga 2 \times 10^6$~K.  
Barring radiative cooling, 
the former
case produces large numbers of 1/4~keV X-rays while
the latter produces both large numbers of 1/4 keV X-rays 
and significant numbers of 3/4 keV X-rays.
When radiative cooling is allowed, the timeframe for bright
emission moves forward and constricts.  The emission rates
also decrease, but some emission is predicted in the
1/4 keV band for $v_z \ge 300$~km~s$^{-1}$ collisions and in the
3/4 keV band for $v_z \ge 400$~km~s$^{-1}$ collisions.
Predictions for \oxyseven\ column densities and intensities
are also provided.
The predicted X-rays originate in the shocked, compressed ambient 
gas, not in the cloud;  the reverse shock is not strong enough to
elevate the cloud's temperature to $\sim 1 \times 10^6$~K.
Although shocked clouds are too cool 
and insufficiently ionized to be X-ray emissive, they may be of interest
for observations of medium and high ions.
%
%

Predictions from the shock-heating scenario
were compared with X-ray count rates observed near Complex C and 
MS30.7-81.4-118, a cloud in the Magellanic Stream.
Moderate density in the ambient medium is required in order for
the shocked-gas to be bright enough to be observed and so it
is possible that high velocity clouds only ``light-up''
upon colliding with moderately dense 
($n_{ISM} = 6.45 \times 10^{-3}$~H~cm$^{-3}$ in Models Ba1, Br1, and C1)
material such as that in the thick disk, Galactic interstellar clouds, 
or material ablated from preceding HVCs.   Furthermore, the
physical conditions of the halo are not fully understood and vary
from location to location.  Our finite set of
simulations cannot reproduce the full spectrum of physical conditions
in the halo, but, instead, provide insights into the possible effects of 
cloud collisions with gas in the halo.

Our simulations of turbulent mixing between cloud and ambient material predict
some X-ray production, but the X-ray count rates are relatively
low unless the clouds move fast enough to also shock heat the
ambient gas.
Excluding cases in which the X-rays originate in shock-heated gas,
our brightest model in
this suite of simulations produces
a 1/4 keV X-ray enhancement of only 
$35 \times 10^{-6}$ counts s$^{-1}$ arcmin$^{-2}$, which is a small
fraction of that seen in the Complex C enhancement 
(e.g., Figure~\ref{fig:complexcmap}).
This comes from a simulation that has a very hot environment 
($T = 3 \times 10^6$~K), moderately dense cloud and 
ambient gas ($0.15$ and $\sim2.5 \times 10^{-4}$~H cm$^{-3}$, respectively), 
and no radiative cooling.
The very high temperature in the environmental gas 
is needed in order to raise that of the mixed gas, such that
some mixed gas has $T \gtrsim 1 \times 10^6$~K.    
The reason why maximizing the output from this
scenario requires moderately dense cloud and ambient
gas is because
the mixed gas falls behind the cloud, where the thermal pressure 
is comparatively low.  
Moderate initial
densities somewhat compensate for the dimming effects of this
low pressure.
Although the interaction zone is not especially bright in X-rays, 
it is rich in high ions.  See \citet{kwak_etal_11} for \carfour, 
\nitfive, and \oxysix\ predictions.

\vspace{1cm}
\noindent
Acknowledgements:

We acknowledge the anonymous referee for his or her 
helpful comments.
The software used in this work was in part developed by the 
DOE-supported
ASC/Alliance Center for Astrophysical Thermonuclear Flashes at the 
University of Chicago. 
The simulations were performed at the Research Computing Center 
(RCC) of the University of Georgia.
We acknowledge financial support from NASA's Astrophysics Theory
and Fundamental Physics Program through grant NNX09AD13G and
support from \chandra's Theory and Modeling Project Program
through grant TM8-9012X.


\end{document}